\documentclass[aps,prd,preprint,floatfix,superscriptaddress,nofootinbib,longbibliography,a4paper]{revtex4-1}
\pdfoutput=1
\usepackage{color}
\usepackage{graphicx}
\usepackage{textcomp}
\usepackage{subfigure}
\usepackage{feynmp}
\usepackage{amsmath,amssymb, array}
\usepackage{enumerate}
\usepackage{slashed}
\usepackage{hhline}
\usepackage{hyperref}
\hypersetup{colorlinks=true,
    citecolor=blue,
	linkcolor=blue,
	filecolor=magenta,      
	urlcolor=blue}
\newcommand{\mathsym}[1]{{}}

\newcommand{\beqa}{\begin{eqnarray}}
\newcommand{\eeqa}{\end{eqnarray}}
\newcommand{\be}{\begin{equation}}
\newcommand{\ee}{\end{equation}}
\newcommand{\ba}{\begin{array}} 
\newcommand{\ea}{\end{array}}

\def\vev#1{\langle #1\rangle}

\begin{document} 
\title{Soft supersymmetry breaking as the sole origin of neutrino masses and lepton number violation}
\bigskip
\author{Anjan S. Joshipura}
\email{anjanjoshipura@gmail.com}
\affiliation{Theoretical Physics Division, Physical Research Laboratory, Navarangpura, Ahmedabad-380009, India}
\author{Ketan M. Patel}
\email{ketan.hep@gmail.com}
\affiliation{Theoretical Physics Division, Physical Research Laboratory, Navarangpura, Ahmedabad-380009, India}

\begin{abstract}
We discuss a scenario in which the supergravity induced soft terms, conventionally used for breaking supersymmetry, also lead to non-zero Majorana neutrino masses. The soft terms lead to the spontaneous violation of the lepton number at the gravitino mass scale $m_{3/2}$ which in turn leads to (i) the Majorana masses of ${\cal O} (m_{3/2})$ for the right-handed neutrinos and (ii) the $R$-parity breaking at the same scale. The former contributes to light neutrino masses through the type I seesaw mechanism, while the latter adds to it through neutrino-neutralino mixing. Both contributions can scale inversely with respect to $m_{3/2}$ given that gaugino and Higgsino masses are also of order $m_{3/2}$. Together, these two contributions adequately explain observed neutrino masses and mixing.  One realization of the scenario also naturally leads to a $\mu$ parameter of ${\cal O} (m_{3/2})$. Despite the lepton number symmetry breaking close to the weak scale, the Majoron in the model exhibits very weak coupling to leptons, satisfying existing constraints on Majoron-lepton interactions. The right-handed neutrinos in the model have a large coupling to Higgsinos. This coupling and the relatively large heavy-light neutrino mixing induced through the seesaw mechanism may lead to the observable signals at colliders in terms of displaced vertices.
\end{abstract}

\maketitle

\section{Introduction}
\label{sec:intro}
Violation of the lepton number, which otherwise is an automatic classical symmetry of the standard model (SM),  provides an appealing avenue to account for small but non-vanishing neutrino masses \cite{Weinberg:1979sa}. It enables neutrinos to become Majorona particles and associates their masses to a new scale in the theory. The exact relation between the neutrino mass scale and a new scale depends on the nature of new physics introduced to violate the lepton number (LN) \cite{Mohapatra:2005wg}. For example, the most popular among these is the type I seesaw mechanism \cite{Minkowski:1977sc,Yanagida:1979as,Mohapatra:1979ia,PhysRevD.22.2227} in which two or more right-handed (RH) neutrinos are introduced and coupled to the SM neutrinos through Yukawa interactions. A Majorana mass $M_{N}$ of RH neutrino $N$ denotes the scale of LN violation in this case. It can be put in by hand or may arise from the spontaneous breaking of a gauge symmetry that includes the LN \cite{Fritzsch:1974nn,GellMann:1980vs}. From the perspective of the light neutrino masses, the RH neutrino mass scale can be as high as $10^{16}$ GeV if they strongly couple to the SM. On the other hand, $M_{N}$ can be as light as of the order of the electroweak scale provided that their couplings with the SM leptons are small.

The other compelling reason to extend the SM comes from the gauge hierarchy problem. The electroweak scale can be stabilised in the presence of some large scale in the theory when the SM is supersymmetrised by replacing it with the minimal supersymmetric standard model (MSSM), see \cite{Martin:1997ns} for a comprehensive review. An efficient implementation of this scheme requires the mass scale of the MSSM fields not very different from the electroweak scale. The existence of all \cite{Dimopoulos:1981zb,Dimopoulos:1981yj,Marciano:1981un} or at least some \cite{Arkani-Hamed:2004ymt,Giudice:2004tc} of the super-partners around the TeV scale also improves the unification of gauge couplings. Theoretically, the mass scale of the MSSM is set by the $\mu$ parameter and the soft supersymmetry (SUSY) breaking scale. The latter is linked to the mass $m_{3/2}$ of the gravitino in theories based on supergravity (see \cite{Lahanas:1986uc} for a review). Like in the case of SM, the MSSM can also be extended to incorporate LN violation as both are independent extensions of the SM. In this case, the scale of LN violation can coexist in the theory independent of $\mu$ and $m_{3/2}$.

This paper aims to point out an interesting scenario in which the soft terms generated in supergravity theory not only break the SUSY but also spontaneously break the LN thereby linking the LN violation scale to the gravitino mass $m_{3/2}$. The amalgamation of the otherwise two distinct scales provides a compelling reason to expect a low-scale LN violation if the SUSY is to resolve the gauge hierarchy problem or it is responsible for the precision unification of the strong and electroweak forces. Soft SUSY breaking terms proportional to $m_{3/2}$ exclusively generate Majorana masses for both the RH and the light neutrinos. The masses of the first are ${\cal O}(m_{3/2})$ while those of the latter are induced through the seesaw mechanism.  We propose a concrete and minimal realisation of the above scenario in this paper. This minimal version is shown to be complete in the sense that it leads to a consistent and predictive description of the neutrino masses and mixing.

It is noteworthy that connections between the mass of RH neutrino(s) and the SUSY breaking scale can also be established through different mechanisms. In the model known as $\mu\nu$MSSM \cite{Lopez-Fogliani:2005vcg,Escudero:2008jg,Ghosh:2008yh,Fidalgo:2009dm,Lopez-Fogliani:2020gzo}, the RH neutrino masses arise through the most general trilinear terms in the superpotential when their superpartners acquire non-vanishing vacuum. The later is linked to the electroweak symmetry breaking scale which is indirectly related to the SUSY breaking through its radiative induction \cite{Ibanez:1982fr}. The LN symmetry is explicitly violated in this model and, therefore, the Dirac and RH neutrino mass terms take the most general form. In another approach \cite{Chun:1995js,Chun:1995bb,Chun:1999kd,Choi:2001cm,Lavignac:2020yld,Joshipura:2021vtf}, LN is spontaneously broken but it's scale is independent of the SUSY breaking scale. However, one of the RH neutrinos is arranged to be a fermionic partner of the Goldstone boson of the global LN symmetry and, therefore, this pair remains massless at the scale of LN violation. Subsequently, the fermionic partner gets a mass of ${\cal O}(m_{3/2})$ when SUSY is broken through supergravity. Even though there is no in-principle connection between the LN breaking scale and $m_{3/2}$ in this kind of setup, the viable neutrino masses in the simplest versions of this framework force these two scales to stay close to each other \cite{Joshipura:2021vtf}. In the framework we propose here, these two scales are inherently and directly interlinked. This connection, along with the presence if LN symmetry in the full theory, is shown to lead to a realistic and more predictive setup for the neutrino masses  compared to previous approaches.

The remainder of the paper is organised as follows. We discuss the basic mechanism leading to a connection between the SUSY breaking and LN violation in the next section. A comprehensive analysis of the light neutrino masses and mixing is presented in section \ref{sec:numass_general}. We compute the couplings of Majoron with the light neutrinos and charged leptons in sections \ref{sec:majoron_neutrino} and \ref{sec:majoron_cl}, respectively, and discuss the phenomenological constraints. In section \ref{sec:RHN_spect}, we discuss the RH neutrino mass spectrum and the possibility of probing them in direct search experiments. The study is concluded in section \ref{sec:disc}.

\section{Lepton number violation from soft SUSY breaking}
\label{sec:model}
In the following, we discuss two different types of frameworks. Both lead to violation of the LN through soft SUSY breaking. However, they are characteristically different in their implication for the neutrino masses. The first scenario assumes a conformal superpotential without any mass scale and the other scenario allows for an explicit $\mu$ parameter.

\subsection{Conformal $W$}
\label{subsec:conformal}
We consider the MSSM extended by three RH neutrino superfields $\hat{N}_\alpha$, ($\alpha=e,\mu,\tau$). To achieve the required scenario, we demand that the superpotential $W$ is conformal, i.e. it does not contain any mass scales. This can be done for example by imposing an $R$ symmetry on $W$ under which all the chiral superfields carry charge $2/3$. This leads to only cubic terms in $W$. All the standard  terms present in the MSSM superpotential, except for the $\mu \hat{H}_1\hat{H}_2$ term, are allowed in this case. The superpotential for the RH neutrinos is completely fixed by imposing an additional global LN-like symmetry under which $(\hat{N}_e,\hat{N}_\mu,\hat{N}_\tau)$ carry charges $(-1,1,0)$. This leads to the following most general superpotential:
\be \label{WRH}
W_{S}=\lambda \hat{N}_e\hat{N}_\mu\hat{ N}_\tau+\frac{\kappa}{3} \hat{N}_\tau^3\,.\ee
Couplings between the MSSM sector and $\hat{N}_\alpha$ also get restricted by the LN symmetry under which the leptonic doublets $\hat{L}_\alpha^\prime$ carry charge $+1$. These are given by
\be \label{WLN}
W_{LN}=\lambda_\alpha\, \hat{L}_\alpha^\prime \hat{H}_2 \hat{N}_e + \lambda_\alpha^\prime\, \hat{L}_\alpha^\prime \hat{H}_1^\prime \hat{\alpha}^c + \eta\, \hat{N}_\tau \hat{H}_1^\prime \hat{H}_2\,, \ee
where $\hat{\alpha}^c$ are the weak singlet charged lepton superfields with lepton number $-1$. The couplings between $\hat{L}_\alpha^\prime$ and $\hat{\alpha}^c$ are chosen flavour diagonal without loss of generality.

Eqs. (\ref{WRH},\ref{WLN}) admit a supersymmetric minimum in which scalar components of all the superfields have a vanishing minimum. All fermions including $\nu_{L \alpha}$, $N_\alpha$ are massless in this limit. These masses are induced by the soft breaking of SUSY. We introduce the following soft terms corresponding to $W_S$:
\be \label{soft1}
V_{\rm soft} = m_{3/2}\left(A_\lambda \lambda \tilde{N}_e \tilde{N}_\mu \tilde{N}_\tau + \frac{1}{3} A_\kappa \kappa \tilde{N}_\tau^3 + {\rm c.c.}\right)+m_{3/2}^2 \sum_{\alpha} a_{N_\alpha}^2 |\tilde{N}_\alpha|^2\,,\ee
where $\tilde{S}$ denotes a Lorentz scalar component of a superfield $\hat{S}$. The neutrino mass generation through these soft SUSY-breaking terms proceeds as follows. Minimization of the scalar potential with these soft terms leads to non-zero vacuum expectation values (VEVs) for $\tilde{N}_\alpha$. These generate the RH neutrino masses which in turn lead to the masses for the left-handed neutrinos through the weak scale type I seesaw mechanism. The non-zero VEVs also spontaneously break  $R$-parity along with the LN, generating additional R-violating contributions to the light neutrino masses. Both these contributions together describe the observed neutrino masses and mixing. We discuss some of the salient features of this neutrino mass generation here and differ the detailed discussion to the next section.

Minimisation of the scalar potential derived from $W_S$ and $V_{\rm soft}$ is carried out and discussed in Appendix \ref{app:minimization}. It leads to the following VEVs:
\be \label{N_vevs}
\langle \tilde{N_\tau} \rangle = \frac{C_\tau}{\lambda} m_{3/2}\,,~~\langle \tilde{N_e} \rangle = \langle \tilde{N_\mu} \rangle = \frac{C_e}{\lambda} m_{3/2}\,,\ee
where $C_\tau$ and $C_e$ are dimensionless parameters determined in general by $\kappa$, $\lambda$ and the soft parameters. The equality in the VEVs of $\tilde{N_e}$ and $\tilde{N_\mu}$ follows from an assumption $a^2_{N_e}=a^2_{N_\mu}$ which makes the full scalar potential invariant under the interchange of $\tilde{N}_e$ and $\tilde{N}_\mu$. In absence of such an assumption, one finds $\langle \tilde{N_e} \rangle \neq \langle \tilde{N_\mu} \rangle$ but with a constraint, $\langle \tilde{N_e} \rangle^2 + \langle \tilde{N_\mu} \rangle^2 = C_e^2 m_{3/2}^2/\lambda^2$. It is noteworthy that all three sneutrinos obtain VEV of the order of the gravitino mass. In this way, the scale of the LN violation is linked to the SUSY breaking scale.

Notably, the $\vev{\tilde{N_\tau}}$ also generates the $\mu$ parameter as in the NMSSM \cite{Ellwanger:2009dp}. But unlike in the NMSSM now the magnitude of $\mu$ is also naturally tied to the gravitino mass:
\be \label{mu}
\mu=-\eta \vev{\tilde{N_\tau}}=-\frac{\eta C_\tau}{\lambda} m_{3/2}\,.\ee
The VEV $\vev{\tilde{N_e}}$ also breaks spontaneously the  $R$-parity \cite{Masiero:1990uj,Barbier:2004ez} and generates the bilinear R-violation from Eq. (\ref{WLN}), described by three parameters:  
\be \label{epsilon}
\epsilon_\alpha\equiv -\lambda_\alpha \vev{\tilde{N_e}}=-\frac{\lambda_\alpha C_e}{\lambda} m_{3/2}\,.\ee

The RH neutrino mass matrix following from Eqs. (\ref{WRH}) and (\ref{N_vevs}) is given, 
in the basis $(N_e,N_\mu,N_\tau)$, by
\be \label{MR}
M_R = m_{3/2} \left(\ba{ccc} 0 & C_\tau & C_e \\
C_\tau & 0 & C_e \\
C_e & C_e & \tilde{C}_\tau \\ \ea \right)\,,\ee
where $\tilde{C}_\tau = \frac{\kappa}{\lambda} C_\tau $. This leads to the three RH  neutrino masses as
\beqa \label{RH_masses}
M_{N_1} &=& C_\tau m_{3/2}\,,\nonumber\\
M_{N_2} &=& \frac{1}{2} C_\tau m_{3/2}\,\left(1 + \frac{\kappa}{\lambda} - \left( \frac{\kappa}{\lambda} - 1\right) \sec2\theta  \right)\,,\nonumber\\
M_{N_3} &=& \frac{1}{2} C_\tau m_{3/2}\,\left(1 + \frac{\kappa}{\lambda} + \left( \frac{\kappa}{\lambda} - 1\right) \sec2\theta  \right)\,,
\eeqa
where
\be \label{t2th}
\tan 2\theta = \frac{-2\sqrt{2}\, C_e}{C_\tau(1 - \kappa/\lambda)}\,.\ee
One also finds
\be \label{detMR}
{\rm Det} M_R = m_{3/2}^3\,C_\tau \left(2 C_e^2 -\frac{\kappa}{\lambda} C_\tau^2\right)\,.\ee
The geometrical mean of three RH neutrino masses is given by $({\rm Det} M_R)^{1/3}$. Typically, ${\rm Det} M_R \sim m_{3/2}^3$ when all the three RH neutrinos have masses of ${\cal O}(m_{3/2})$. ${\rm Det} M_R \ll m_{3/2}^3$  implies one or more RH neutrinos with a mass smaller than $m_{3/2}$.

The above $M_R$ together with the Dirac mass matrix following from Eq. (\ref{WLN}) leads to the seesaw contribution to the light neutrino masses as
\be \label{mnuseesaw}
({\cal M}^{\rm SS}_\nu)_{\alpha\beta}=\tilde{m}_0\, \lambda_\alpha\lambda_\beta~.\ee
where 
\be \label{tildem0}
\tilde{m}_0=\frac{v_2^2}{m_{3/2}}\frac{\lambda C_e^2}{C_\tau (2\lambda  C_e^2 -\kappa C_\tau^2)} = \frac{v_2^2 C_e^2}{m_{3/2}} \left(\frac{m_{3/2}^3}{{\rm Det} M_R}\right)\,.\ee
The neutrino masses are determined by the Dirac neutrino Yukawa couplings and the inverse of $m_{3/2}$. Small (high) values of the former (latter) can result in the required suppression in the neutrino masses as it happens in the usual seesaw case. The neutrino mass thus would be automatically suppressed in models with high-scale SUSY breaking.

Interestingly, Eqs. (\ref{mnuseesaw},\ref{tildem0}) offer an alternative way of lowering the seesaw contribution to neutrino masses even in low-scale SUSY models. This can be done if the parameter $\lambda/\kappa$ in Eq. (\ref{WRH}) is taken very small\footnote{This can be done in a technically natural way since the limit $\lambda \rightarrow 0$ increases the symmetry of $W_S$.}. This breaks LN at a scale higher than $m_{3/2}$, see Eq. (\ref{N_vevs}), and results in a small $ \tilde{m}_0$. This way of suppressing neutrino masses can be interpreted in terms of the inverse seesaw mechanism often used for suppressing neutrino masses in the weak or TeV scale seesaw models \cite{Mohapatra:1986bd,Mohapatra:1986aw,Nandi:1985uh}. It follows from Eq. (\ref{RH_masses}) that one of the RH neutrinos, namely $M_{N_3}$, becomes heavier compared to the other two if $\lambda/\kappa \ll 1$.

Integrating out $N_\tau$ results in a $5\times 5$ mass matrix, written in the basis $(\nu^\prime_\alpha, N_e, N_\mu)$,
\be \label{inverse}
{\cal M}_\nu^{\rm IS}=\left(\ba{ccc}
0_{3\times3}& m_D&0\\
m_D^T&\mu_S&M_D\\
0&M_D&\mu_S\\
\ea\right)~,\ee
where $M_D=m_{3/2} C_\tau$ and $m_D$ is $3\times 1$ matrix of the Yukawa couplings $\lambda_\alpha$ multiplied by the Higgs VEV $v_2$. The matrix ${\cal M}_\nu^{\rm IS}$ has a form typically used in the inverse seesaw with 
\be \label{muS}
\mu_S \equiv -\frac{\lambda}{\kappa} \frac{C_e^2 m_{3/2}}{C_\tau}\,, \ee
representing the small LN breaking parameter suppressed by $\lambda/\kappa$. For $M_D \gg m_D$,  Eq. (\ref{inverse}) leads to the light neutrino mass matrix
\be \label{light_nu_inverse}
\frac{\mu_S v_2^2}{m_{3/2}^2 C_\tau^2}\,\lambda_\alpha\lambda_\beta\, \ee
which coincides with the seesaw result, Eq. (\ref{mnuseesaw}), for $\lambda/\kappa \ll 1$.

It follows from Eq. (\ref{WLN}) that only one combination of the neutrino fields $\lambda_\alpha \nu_\alpha'$
couples to $N_e$ and obtains its mass through the canonical seesaw mechanism. An independent linear combination of the neutrino fields obtains its mass from the $R$-parity violation induced by the soft terms. The $R$-violation is characterised by $\epsilon_\alpha$ given in Eq. (\ref{epsilon}). They give rise to the VEV for the left-handed sneutrino leading to the mixing of neutrinos with neutralinos. This mixing generates mass for the other combination of neutrinos.  The $R$-parity violating parameters $\epsilon_\alpha$ increase with the gravitino mass. So naively, one would expect that this contribution to the neutrino masses would increase with $m_{3/2}$ in contrast to the seesaw contribution. However, this does not happen and both contributions can be inversely proportional to $m_{3/2}$ in a typical situation as we show below.

The $R$-parity violating contribution is specified in the rotated basis in which bilinear $R$-parity violation is rotated away from the superpotential \cite{Joshipura:2002fc}. These unprimed bases are defined as
\beqa \label{basis}
\hat{H}_1&=& \hat{H}_1^\prime+\frac{\epsilon_\alpha}{\mu} \hat{L}_\alpha^\prime+{\cal O}\left(\frac{\epsilon_\alpha^2}{\mu^2}\right)\,,\nonumber\\
\hat{L}_\alpha &=& \hat{L}_\alpha^\prime-\frac{\epsilon_\alpha}{\mu} \hat{H}_1^\prime + {\cal O}\left(\frac{\epsilon_\alpha^2}{\mu^2}\right)\,,\eeqa
where $\hat{L}_\alpha^\prime$ and $\hat{H}_1^\prime$ are the original fields present in the superpotential Eq. (\ref{WLN}). The $R$-parity  violating light neutrino mass matrix is obtained as (see the next section for the derivation)
\be \label{mnurp}
\left({\cal M}_\nu^R \right)_{\alpha\beta} = A_0\, \omega_\alpha \omega_\beta\,,\ee
with
\be \label{A0}
A_0 = - \frac{\mu (g^2 M_1 + g^{\prime 2} M_2)}{2 (M_1 M_2 \mu - v_1 v_2 (g^2 M_1 + g^{\prime 2} M_2))}\,.\ee
Here $M_{1,2}$ are gaugino masses and $v_{1,2}$ are the VEVs of the Higgs fields $\tilde{H}_{1,2}$.

In Eq. (\ref{mnurp}), $\omega_\alpha \equiv \langle \nu_\alpha \rangle$ denote the VEVs of the scalar superpartners of the electrically neutral component of $\hat{L}_\alpha$. They are determined in terms of the soft parameters by minimising the relevant scalar potential. Writing 
\be \label{k_def}
\omega_\alpha=k_\alpha \epsilon_\alpha\,, \ee 
we find in this case
\be \label{kalpha1}
k_\alpha \simeq \frac{v_1}{\mu} \left(\frac{a_{H_1}^2 - a_{L_\alpha}^2}{a_{L_\alpha}^2} \right) + \frac{v_1 \tan\beta}{m_{3/2}} \left( \frac{A_\eta - A_\alpha}{a_{L_\alpha}^2}\right) \,,\ee
where $\tan \beta=v_2/v_1$. The details of the derivation of the above can be found in Appendix \ref{app:k_calc}. If one assumes universal boundary conditions at a high scale then the differences in soft parameters appearing in $k_\alpha$ are generated at the low scale through renormalization effects and are small.  In particular, the first term which gives a dominant contribution  to $k_\alpha$ is typically given by 
\be \label{k_magnitude} 
\frac{a_{H_1}^2-a_{L_\alpha}^2}{a_{L_\alpha}^2} \sim \frac{3 y_b^2}{4 \pi^2}\, \log\frac{M_X}{M_Z}\sim {\cal O} \left(10^{-3}\right)\,.\ee
Here, $y_b$ is the bottom quark Yukawa coupling and $M_X$ is the scale where the universality of the soft terms is imposed.

It is useful to eliminate Yukawa couplings and consider the ratio of the contributions ${\cal M}_\nu^R$ and ${\cal M}_\nu^{\rm SS}$. We find
\be \label{ratio}
\xi_0 \equiv \frac{({\cal M}_\nu^R)_{33}}{({\cal M}^{\rm SS}_\nu)_{33}}=\frac{A_0 k_\tau^2 m_{3/2}^3}{\lambda^2 v_2^2}\, \left(\frac{{\rm Det} M_R}{m_{3/2}^3}\right)\,.\ee
Following Eqs. (\ref{kalpha1},\ref{k_magnitude}), $k_\tau$ can be approximated as
\be \label{ktau1}
 k_\tau\approx 10^{-3}\, \frac{v_1}{\mu}~.\ee
This implies 
\be \label{r1num}
\xi_0 \approx \frac{A_0 m_{3/2}^3}{\mu^2 \tan^2\beta} \left(\frac{10^{-6}}{\lambda^2}\right)\,  \left(\frac{{\rm Det} M_R}{m_{3/2}^3}\right)\,.\ee

Since $\xi_0$ is related to the two non-zero neutrino masses, it cannot be too small or large. The detailed fit to neutrino masses and mixing presented in the next section show that the magnitude of $\xi_0$ is of order unity. Requiring $\xi_0\sim 1$ has implications for  the allowed $m_{3/2}$ and the SUSY spectrum. Considering $M_{N_i} \sim m_{3/2}$, we note that:
\begin{itemize}
\item If gaugino masses and the $\mu$ parameter are similar to $m_{3/2}$ then $\xi_0$ is independent of $m_{3/2}$ since $A_0$ is inversely proportional to the gaugino masses. This implies that like the seesaw contribution, the $R$-parity violating contribution to the neutrino masses is also inversely proportional to $m_{3/2}$. Therefore, the neutrino masses can be suppressed without requiring too small Yukawa couplings as in the high-scale seesaw models. This scenario would conflict with the gauge coupling unification which requires gauginos at the electroweak scale.
\item If $\mu$ is held fixed around electroweak scale by choosing $\eta$ appropriately and the gaugino masses are also assumed to be of the order of $\mu$ ($m_{3/2}$) then the $\xi_0$ is proportional to $m_{3/2}^3/\mu^3$ ($m_{3/2}^2/\mu^2$). Then the requirement, $\xi_0 \sim 1$, implies an upper bound on $m_{3/2}/\mu$. We will explore this point more quantitatively in the next section. 
\item $\lambda$ cannot be arbitrarily small and can be around $10^{-2}$ if $\mu$ parameter and the gaugino masses are similar to $m_{3/2}$. Thus, $\lambda$ can provide only a mild suppression in the neutrino masses through inverse seesaw.
\end{itemize}
Some of the above features change when $W$ is not conformal and contains an independent scale as a bare $\mu$ term. We discuss this below.

\subsection{$W$ with a bare $\mu$-term}
An explicit $\mu$ parameter can exist if $R$ symmetry used to obtain cubic potential is not insisted upon. This has a significant effect on neutrino masses which we now discuss. Consider the same superpotential with an added bare $\mu$-term, namely
\beqa \label{W}
W & = & W_S + W_{LN} - \mu_0\, \hat{H}^\prime_1 \hat{H}_2\,.\eeqa
The effective $\mu$ parameter is now obtained as
\be \label{newmu}
\mu=-\eta\vev{\tilde{N}_\tau}+\mu_0\,.\ee
We can altogether omit the $\eta$-dependent term. In this case, $\mu=\mu_0$ is not tied to the gravitino mass and represents an independent scale. This can be technically done by imposing a discrete $Z_3$ symmetry under which $\hat{\alpha}^{c} \rightarrow \omega^2 \hat{\alpha}^{c}$, $\hat{N}_\alpha \rightarrow \omega^2 \hat{N}_\alpha$ and $\hat{L}'_\alpha \rightarrow \omega \hat{L}'_\alpha$ with $\omega^3=1$.

The  $\mu_0$ term leads to an additional soft term $B_{\mu_0} \mu_0 \tilde{H}^\prime_1 \tilde{ H}_2$ in the scalar potential.  The $k_\alpha$ of Eq. (\ref{kalpha1}) is now replaced by a more general expression
\be \label{kalpha2}
k_\alpha \simeq \frac{v_1}{\mu} \left( \frac{a_{H_1}^2 - a_{L_\alpha}^2}{a_{L_\alpha}^2}\right) + \frac{v_1 \tan\beta}{m_{3/2}} \left( \frac{A_\eta - A_\alpha}{a_{L_\alpha}^2}\right) + \frac{\mu_0}{\mu}\frac{v_1 \tan\beta}{m_{3/2}} \left( \frac{B_{\mu_0} - A_\alpha}{a_{L_\alpha}^2}\right)\,.\ee
The above is obtained using Eq. (\ref{k_calculated}) derived explicitly in Appendix \ref{app:k_calc}.  Unlike in the previous case,  $k_\alpha$ is non-zero even at a high scale since generally $B$ and $A$ parameters are not equal in supergravity-induced soft terms. The last term, therefore, dominates  in Eq. (\ref{kalpha2}). This increases the $R$-parity violating contribution by a factor of around $10^{6}$. Since the last term in $k_\alpha$ dominates,
\be \label{ktau2}
k_\tau \approx \frac{\mu_0}{\mu} \frac{v_2}{m_{3/2}}\,.\ee
The ratio $\xi_0$ defined in Eq. (\ref{ratio}) is now given by
\be \label{r2num}
\xi_0 \approx A_0 m_{3/2} \left(\frac{\mu_0}{\mu}\right)^2 \left(\frac{1}{\lambda^2}\right)\,  \left(\frac{{\rm Det} M_R}{m_{3/2}^3}\right)\,.\ee
This suggests that $\lambda$ cannot be chosen small ruling out the possibility of an inverse seesaw mechanism entirely in this case. Also since  $A_0$ is inversely proportional to the gaugino mass, one requires this mass to be ${\cal O}(m_{3/2})$ and $\mu_0\sim \mu$ to have  $\xi_0$ of  order one.

We have neglected in the above discussions one more contribution to the light neutrino masses which arise from the mixing of the RH neutrinos with Higgsino originating from Eq. (\ref{WLN}). This gives a much smaller contribution compared to the ones already discussed. Inclusion of this requires consideration of the full mass matrix of neutral fermions in the model which we discuss in the following section.

\section{Light neutrino masses and mixing}
\label{sec:numass_general}
The model has 10 neutral fermions denoted as $\psi_0 = (\nu_\alpha',N_\alpha, -i \lambda_1, -i \lambda_2, \tilde{h}_1^0,\tilde{h}_2^0)^T$. The RH VEVs generated through soft terms cause mixing among them as mentioned in the earlier section. The full  mixing generated by the superpotential in Eq. (\ref{W}) and gauge interactions can be written as:  
\be \label{Lneut}
-{\cal L}_{\cal N} = \frac{1}{2}\, \psi_0^T\, {\cal M}_{\cal N}\, \psi_0\,+\,{\rm h.c.}\,, \ee
where ${\cal M}_{\cal N}$  is a $10 \times 10$ matrix parametrized as
\be \label{Mneut}
{\cal M}_{\cal N} = \left(\ba{ccc} (0)_{3 \times 3} & (m_{\nu N})_{3\times 3} & (m_{\nu \chi})_{3 \times 4} \\
                      m_{\nu N}^T & (M_R)_{3 \times 3} & (m_{ N\chi})_{3 \times 4} \\
                      m_{\nu\chi}^T & m_{ N\chi}^T & (M_\chi)_{4 \times 4}  \ea \right)\,.\ee
Here,
\beqa \label{mnuN}
m_{\nu N} & = & v_2\left(\ba{ccc} \lambda_e & 0 & 0\\ \lambda_\mu & 0 & 0 \\ \lambda_\tau & 0 & 0 \ea \right)\,,~~
m_{\nu \chi} = \left(\ba{cccc} -\frac{1}{\sqrt{2}}g^\prime \omega^\prime_e & \frac{1}{\sqrt{2}}g \omega^\prime_e & 0 & -\epsilon_e \\
-\frac{1}{\sqrt{2}}g^\prime \omega^\prime_\mu & \frac{1}{\sqrt{2}}g \omega^\prime_\mu & 0 & -\epsilon_\mu \\
-\frac{1}{\sqrt{2}}g^\prime \omega^\prime_\tau & \frac{1}{\sqrt{2}}g \omega^\prime_\tau & 0 & -\epsilon_\tau \ea \right)\,. \eeqa
The $M_R$ is already given in Eq. (\ref{MR}) while 
\beqa \label{}
m_{N \chi} &=& \left(\ba{cccc} 0 & 0 & 0 & \lambda_\alpha \omega_\alpha^\prime \\
0 & 0 & 0 & 0  \\ 0 & 0 & \eta v_2 & \eta v_1 \ea \right)\,,~~
M_{\chi} = \left( 
\ba{cccc}
 M_1 & 0 & -\frac{1}{\sqrt{2}} g^\prime v_1 &  \frac{1}{\sqrt{2}}g^\prime v_2 \\
 0 & M_2 & \frac{1}{\sqrt{2}} g v_1 & - \frac{1}{\sqrt{2}}g v_2 \\
-\frac{1}{\sqrt{2}} g^\prime v_1 & \frac{1}{\sqrt{2}} g v_1 & 0 & -\mu  \\
 \frac{1}{\sqrt{2}} g^\prime v_2 & -\frac{1}{\sqrt{2}} g v_2 & -\mu  & 0 \\ \ea \right)\,.\eeqa
$M_\chi$ is the standard neutralino mass matrix of the MSSM \cite{Martin:1997ns}. $R$ parity-violating couplings mix the neutralinos with the SM neutrinos. This mixing is denoted as $m_{\nu\chi}$. Specifically, Higgsino-neutrino mixing is given by $\epsilon_\alpha$ and gaugino mixing is determined by the sneutrino VEVs $\omega_\alpha^\prime \equiv \vev{\nu_\alpha'}$. $m_{N\chi}$ denotes the mixing of the singlet states $N_\alpha$ with neutralinos following from Eq. (\ref{W}). $m_{\nu N}$ represents $U(1)$ invariant Dirac mass term between neutrinos and singlet states. The above ${\cal M}_{\cal N}$ describes the general case with explicit $\mu_0$ term present. The conformal case is obtained by putting $\mu_0=0$ in $\mu$ appearing in ${\cal M}_{\cal N}$.

\subsection{Analytical form of effective neutrino mass matrix}
The tree-level light neutrino mass matrix can be obtained  from Eq. (\ref{Mneut}) in the seesaw approximation as
\be \label{mnu}
m_\nu = -m_{3\times 7}\, {\cal M}_{7\times 7}^{-1}\,(m_{3\times 7})^T\,,\ee
where $m_{3\times 7} = (m_{\nu N}, m_{\nu N}, m_{\nu \chi})$ and ${\cal M}_{7 \times 7}$ is the lower-right $7 \times 7$ block of the full matrix given in Eq. (\ref{Mneut}). Neglecting $\lambda_\alpha \omega_\alpha^\prime$ term in $m_{N \chi}$ and performing some algebraic simplification, we get
\beqa \label{mnu_ana}
m_\nu &=& \tilde{m} \left(a\,{\cal A} + b\, {\cal B} + c\,{\cal C} \right) \,.\eeqa
Here,
\be \label{mtilde}
\tilde{m} = \frac{D_4 v_2^2 C_e^2 m_{3/2}^2}{D_7}\,,\ee
with
\be \label{D4}
D_4={\rm Det}\,M_\chi = -\mu(M_1 M_2 \mu - (g^2 M_1 + g^{\prime 2} M_2) v_1 v_2)\,, \ee
and 
\be \label{d7}
D_7= D_4 C_\tau m_{3/2}^3 \left(\frac{2 \lambda C_e^2 - \kappa C_\tau^2}{\lambda} - \frac{\eta^2 C_\tau \zeta}{m_{3/2}}\right)\,,\ee
is determinant of $7 \times 7$ matrix ${\cal M}_{7\times 7}$ with
\be \label{zeta}
\zeta = \frac{v^4 (g^2 M_1 + g^{\prime 2}M_2)-4 v_1 v_2 \mu M_1 M_2}{2 D_4}\,.\ee
Substituting Eq. (\ref{d7}) in Eq. (\ref{mtilde}) and using Eq. (\ref{tildem0}), we find
\be \label{m-m0}
\tilde{m} = \frac{\tilde{m}_0}{1-\frac{\lambda \eta^2 C_\tau}{(2 \lambda C_e^2 - \kappa C_\tau^2)} \frac{\zeta}{m_{3/2}}}\,.\ee
Since $\zeta/m_{3/2} \ll 1 $ for $v_{1,2} \ll m_{3/2}$, the denominator provides only a small correction to the equality, $\tilde{m} = \tilde{m}_0$.

The matrices in Eq. (\ref{mnu_ana}) are obtained as 
\be \label{AB}
{\cal A} =\left( \ba{ccc} \lambda_e^2 & \lambda_e \lambda_\mu & \lambda_e \lambda_\tau\\
\lambda_e \lambda_\mu & \lambda_\mu^2 &  \lambda_\mu \lambda_\tau\\
\lambda_e \lambda_\tau & \lambda_\mu \lambda_\tau & \lambda_\tau^2 \\ \ea \right)\,,
~{\cal B} = \left( \ba{ccc} \omega_e^2 & \omega_e \omega_\mu & \omega_e \omega_\tau\\
\omega_e \omega_\mu & \omega_\mu^2 &  \omega_\mu \omega_\tau\\
\omega_e \omega_\tau & \omega_\mu \omega_\tau & \omega_\tau^2 \\ \ea \right)\,,
\ee
\be \label{C}
{\cal C} = \left( \ba{ccc} 2 \lambda_e \omega_e & \lambda_e \omega_\mu + \lambda_\mu \omega_e & \lambda_e \omega_\tau + \lambda_\tau \omega_e\\
\lambda_e \omega_\mu + \lambda_\mu \omega_e & 2 \lambda_\mu \omega_\mu &  \lambda_\mu \omega_\tau + \lambda_\tau \omega_\mu \\
\lambda_e \omega_\tau + \lambda_\tau \omega_e &  \lambda_\mu \omega_\tau + \lambda_\tau \omega_\mu & 2 \lambda_\tau \omega_\tau \\ \ea \right)\,. \ee
In the above, 
\be \label{omega}
\omega_\alpha =  \omega^\prime_\alpha - \frac{\epsilon_\alpha}{\mu} v_1 \,,\ee
denote the VEVs of sneutrino in the basis in which the bilinear term is rotated away from the superpotential, see Eq. (\ref{basis}). Finally, the dimensionless coefficients in Eq. (\ref{mnu_ana})  are obtained as
\beqa \label{abc}
a &=& \left(1 - \frac{\eta C_\tau m_{3/2}}{\lambda \mu}\right)^2 = \left(2-\frac{\mu_0}{\mu} \right)^2\,, \nonumber\\
b &=& - \frac{A_0 C_\tau^2 \tilde{C}_\tau m_{3/2}}{C_e^2 v_2^2} \left(1 - 2 \frac{C_e^2}{C_\tau \tilde{C}_\tau} + 2  \frac{\eta^2 v_1 v_2}{\tilde{C}_\tau \mu m_{3/2}}\right)\,, \nonumber \\
c &=&\frac{A_0 \eta C_\tau (v_2^2- v_1^2)}{C_e v_2 \mu} \left(1 - \frac{\eta C_\tau m_{3/2}}{\lambda \mu}\right) = \frac{A_0 \lambda (v_2- v_1^2/v_2)}{C_e m_{3/2}} \left(\frac{\mu_0}{\mu}-1\right)\left(2- \frac{\mu_0}{\mu}\right)\,,\eeqa
and $A_0$ is defined in Eq. (\ref{A0}).

The structures of the first two terms in Eq. (\ref{mnu_ana}) coincide with the seesaw and $R$ parity violating terms obtained in the earlier section. The last term is  generated by the mixing among the RH neutrino $N_\tau$ and Higgsino coming from Eq. (\ref{W}). This mixing also modifies the overall coefficients of the seesaw and $R$-violating interactions compared to the simplified case considered earlier. Using $\omega_\alpha = k_\alpha \epsilon_\alpha $, the neutrino mass matrix can be further simplified to the following form 
\be \label{mnu_ana_simp}
m_\nu =  m \left(
\begin{array}{ccc}
r_e^2 \left(1+\xi  s_e^2+2 \delta  s_e \right) & r_e r_{\mu } \left(1+\delta  s_{\mu }+s_e \left(\delta +\xi  s_{\mu }\right)\right) & r_e \left(1+\delta +(\delta +\xi ) s_e\right) \\
r_e r_{\mu } \left(1+\delta  s_{\mu }+s_e \left(\delta +\xi  s_{\mu }\right)\right) & r_{\mu }^2 \left(1+\xi  s_\mu^2+2 \delta  s_\mu \right) & r_{\mu } \left(1+\delta +(\delta +\xi ) s_{\mu }\right) \\
r_e \left(1+\delta +(\delta +\xi ) s_e\right) & r_{\mu } \left(1+\delta +(\delta +\xi ) s_{\mu }\right) & 1+ \xi  + 2 \delta\\ \end{array} \right),\ee
where we have define $r_{e,\mu}=\frac{\lambda_{e,\mu}}{\lambda_\tau}$, $k_\tau \equiv k$ and $s_{e,\mu} = \frac{k_{e, \mu}}{k}$. The parameters $k_\alpha$ are given by Eqs. (\ref{kalpha1}) or Eq. (\ref{kalpha2}) depending on the choice of $\mu$-term in the model.
Also, 
\beqa \label{m}
m &=&\tilde{m}\lambda_\tau^2 a = \tilde{m}_0 \lambda_\tau^2 a \left(1-\frac{\lambda \eta^2 C_\tau}{(2 \lambda C_e^2 - \kappa C_\tau^2)} \frac{\zeta}{m_{3/2}}\right)^{-1} \,,\nonumber\\
\xi &=& \frac{\xi_0}{a}\left(1-\frac{2 \lambda \eta^2 v_1v_2C_\tau}{m_{3/2}\mu(2 \lambda C_e^2 - \kappa C_\tau^2)}\right) \,,\nonumber \\
\delta& =& -\frac{c}{a}\frac{k C_e m_{3/2}}{\lambda}\,.\eeqa
Here, $\tilde{m}_0$ and $\xi_0$ are defined in Eqs. (\ref{tildem0},\ref{ratio}), respectively. $a \tilde{m}_0$  denotes  the overall scale of the seesaw contribution. The $\xi$ corresponds to the ratio of the $R$-parity violating and seesaw terms. $\tilde{m}$ and $\xi$ respectively reduce to $\tilde{m}_0$ and $\xi_0$ of the earalier section when $\eta = 0$. In the following, we shall only be working with the tree-level neutrino mass matrix and will neglect the other sub-dominant radiative contributions. These include corrections to the seesaw mass matrix \cite{Grimus:2018rte} and corrections generated by the effective trilinear R parity violating  terms present in the rotated basis \cite{Barbier:2004ez}.  The tree-level mass matrix considered here is by itself sufficient to reproduce the neutrino mass spectrum and the radiative corrections are not expected to change the results significantly.

The determinant of $m_\nu$ is vanishing. It predicts a massless neutrino. The eigenvector corresponding to the massless state can be determined using $\sum_j(m_\nu)_{ij}\,u_j = 0$. This gives 
\be \label{ev_m0}
\left(u_1,u_2,u_3\right) = \pm \frac{1}{\sqrt{N}}\,\left(r_\mu (s_\mu-1),\, - r_e(s_e-1),\, r_e r_\mu (s_e-s_\mu)\right)\,.\ee
In the diagonal basis of the charged leptons, the above is identified with the first (third) column of the leptonic mixing matrix for normal (inverted) hierarchy in the neutrino masses. The observed neutrino mixing angles then necessarily require $s_e \neq 1$, $s_\mu \neq 1$ as well as $s_e \neq s_\mu$. This requires sizeable flavour violations in the soft terms, a feature which was noticed also earlier in \cite{Joshipura:2002fc,Romao:1999up,Hirsch:2000ef} in the context of different frameworks.

\subsection{Example solutions}
We now demonstrate that the effective neutrino mass matrix, $m_\nu$ as obtained in Eq. (\ref{mnu_ana_simp}), can account for the realistic neutrino masses and mixing observables. As mentioned earlier, the charged lepton Yukawa couplings can be chosen real and diagonal without loss of generality. Moreover, the parameters $\lambda$, $\lambda_\alpha$, $\eta$ and $\mu_0$ appearing in the superpotential can be made real through redefinitions of various superfields. This along with an assumption for the real VEVs of $N_\alpha$ and $\nu_\alpha$ lead to real $r_{e,\mu}$, $s_{e,\mu}$ and $\delta$ in Eq. (\ref{mnu_ana_simp}). The phase of $m$ is also unphysical. Altogether, these six real parameters and complex $\xi$ are required to reproduce the realistic values of six observables, namely the solar and atmospheric neutrino masses, three mixing angles and the Dirac CP phase.
\begin{table}[t]
\begin{center}
\begin{tabular}{ccccc} 
\hline
\hline
~~~~Parameters~~~~ & ~~~~NH1~~~~  & ~~~~NH2~~~~ & ~~~~NH3~~~~ & ~~~~NH4~~~~\\
\hline
$m$ [eV] & 1.0 & 1.0  & 0.01 &  0.01 \\
$\xi$ & -0.9831 & 1.0216  & -2.6937 &  -1.7155 \\
$\delta$ &  0 & -0.9983 & 0 & -0.4891 \\
$r_e$ & 0.0782 & -0.1196 & 1.387 & -1.3511\\
$r_\mu$ & -0.5964 & 0.9562 & -3.2619 & 3.1967\\
$s_e$ & 1.1216 & 0.5015 & 0.5859 & 0.501\\
$s_\mu$ & 0.971 & 0.8866& 0.6797 & 0.6163\\
\hline
~~~~Parameters~~~~& ~~~~NH5~~~~  & ~~~~NH6~~~~ & ~~~~NH7~~~~ & ~~~~NH8~~~~ \\
\hline
$m$ [eV] & 1.0 & 1.0  & 0.01 &  0.01 \\
$|\xi|$ & 0.9841 & 0.5857  & 2.704 &  1.8462 \\
${\rm Arg}[\xi]$ & -3.1349 & -0.01  & -3.096 &  -3.0176 \\
$\delta$ &  0 & -0.7847& 0 & -0.4372 \\
$r_e$ & -0.0814 & 0.1684& 1.3628 & -1.3123\\
$r_\mu$ & -0.6049 & -0.4458 & 3.2386 & -3.1601\\
$s_e$ & 1.1158 & 1.3175 & 0.5813 & 0.5004\\
$s_\mu$ & 0.9714 & 0.7802 & 0.678 & 0.6197\\
\hline
\hline
\end{tabular}
\end{center}
\caption{Example solutions of the parameters of the effective neutrino mass matrix which reproduces the neutrino masses and mixing parameters within $1\sigma$ assuming normal hierarchy. The first four solutions correspond to the CP conserving case.}
\label{tab:sol_NH}
\end{table}
\begin{table}[t]
\begin{center}
\begin{tabular}{ccccc} 
\hline
\hline
~~~~Parameters~~~~& ~~~~IH1~~~~ & ~~~~IH2~~~~ & ~~~~IH3~~~~ & ~~~~IH4~~~~ \\
\hline
$m$ [eV] & 1.0 & 1.0  & 0.01 &  0.01 \\
$\xi$ & -0.982 & -0.5409  & -0.4514 &  -0.347 \\
$\delta$ &  0 & -0.2272 & 0 & -0.1066 \\
$r_e$ & -0.2517 &  -0.2393 & -11.2255 & 14.6181\\
$r_\mu$ & -0.9077 & 0.8048 & 3.0287 & 3.7629\\
$s_e$ & 1.1493 & 1.2026 & 1.5 & 1.4252\\
$s_\mu$ & 1.0082 & 0.9881 & 1.3626 & 1.3282\\
\hline
~~~~Parameters~~~~& ~~~~IH5~~~~  & ~~~~IH6~~~~ & ~~~~IH7~~~~ & ~~~~IH8~~~~ \\
\hline
$m$ [eV] & 1.0 & 1.0  & 0.01 &  0.01 \\
$|\xi|$ & 0.9826 & 0.0836  & 3.0314 &  2.8087 \\
${\rm Arg}[\xi]$  & -3.1412 & -0.0039  & -3.1386 &  -3.1384 \\
$\delta$ &  0 & -0.5321 & 0 & -0.0918 \\
$r_e$ &-0.5623& -0.6534& 4.3557 & 4.4436\\
$r_\mu$ & -0.743 & 0.725 & -1.7175 & 1.735\\
$s_e$ & 1.0669 & 1.1278 & 0.5083 & 0.5\\
$s_\mu$ & 0.99 & 0.9772 & 0.7529 & 0.7463\\
\hline
\hline
\end{tabular}
\end{center}
\caption{Four example solutions of the parameters of the effective neutrino mass matrix which reproduces the neutrino masses and mixing parameters within $1\sigma$ assuming inverted hierarchy. The first four solutions correspond to the CP conserving case.}
\label{tab:sol_IH}
\end{table}

Following the usual $\chi^2$ optimization technique \cite{Joshipura:2021vtf}, we find several example solutions for the parameters in $m_\nu$ which lead to the viable neutrino spectrum. First, we consider real $\xi$ which leads to the vanishing CP phase, $\delta_{\rm CP}$. The latter is not directly observed and the present fits to the neutrino oscillation data \cite{Capozzi:2017ipn,deSalas:2020pgw,Esteban:2020cvm} allows $\delta_{\rm CP}=0$ within $\pm 3\sigma$. We find solutions for two example values of $m$ and also consider a possibility in which the parameter $\delta$ can be zero. The solutions obtained in this way are displayed as the first four solutions in Table \ref{tab:sol_NH} (\ref{tab:sol_IH}) for normal (inverted) hierarchy in the neutrino masses. Subsequently, we also consider complex $\xi$ to account for the Dirac CP phase. These solutions are listed as NH5 to NH8 in the case of normal and IH5 to IH8 in the case of inverted hierarchy in the respective tables. All the NH (IH) solutions give $\Delta m_{21}^2 = 7.41 \times 10^{-5}\,{\rm eV}^2$, $\Delta m_{31}^2 = 2.511 \times 10^{-3}\,{\rm eV}^2$ ($\Delta m_{23}^2 = 2.498 \times 10^{-3}\,{\rm eV}^2$), $\sin^2\theta_{12} = 0.303$, $\sin^2\theta_{23}=0.572$ ($\sin^2\theta_{23}=0.578$) and $\sin^2\theta_{13}=0.02203$ ($\sin^2\theta_{13}=0.02219$). Solutions with complex $\xi$ give $\sin\delta_{\rm CP} = -0.3$ ($\sin\delta_{\rm CP} = -0.96$) for the NH (IH) case. All these values correspond to the central values as obtained by the latest fit to the neutrino oscillation data (see, NuFIT 5.2) in \cite{Esteban:2020cvm}.

The noteworthy feature of these fits is that $|\xi| \sim 1$ implying that the scales associated with both the $R$-violating contribution and the seesaw terms are similar in magnitudes. This can be used to draw important conclusions on the basic parameters of the model. The exact expression of $\xi$, using Eqs. (\ref{abc},\ref{detMR}), can be rewritten as
\be \label{xi_simpl}
\xi = \frac{A_0 k^2}{\lambda^2 v_2^2} {\rm Det} M_R \left(2-\frac{\mu_0}{\mu} \right)^{-2} \left(1-\frac{2 \lambda^2 v_1 v_2 \mu}{{\rm Det} M_R} \left(1-\frac{\mu_0}{\mu} \right)^2 \right)\,.\ee
Scenarios in which the common gaugino mass scale, $M_1 \simeq M_2 \simeq \mu \simeq \mu_0 \equiv M_g$, is arranged to stay close to the electroweak scale and $m_{3/2} \gg v$, then leads to
\be \label{xi_simpl_2}
\xi \simeq \frac{m_{3/2}}{2 M_g}\, \frac{1}{\lambda^2}\,\left(\frac{{\rm Det} M_R}{m_{3/2}^3}\right)\,.\ee
While deriving the above, we have used the expression Eq. (\ref{ktau2}) for the $k$ parameter. The fact that the value of $|\xi|$   cannot  be arbitrarily large, as indicated by the fits, implies that the gaugino mass scale cannot be much smaller than $m_{3/2}$ if ${\rm Det} M_R \sim m_{3/2}^3$. In other words, it is not possible to arrange gauginos at the electroweak scale and all the RH neutrinos at an arbitrarily high scale. The ratio $\frac{{\rm Det} M_R}{m_{3/2}^3}$ is determined by the soft SUSY breaking parameters and $\frac{\kappa}{\lambda}$, see Eq. (\ref{detMR}). In supergravity models, the soft parameters are naturally of order unity and hence $\frac{{\rm Det} M_R}{m_{3/2}^3} \sim {\cal O}(1)$. However, it can be suppressed in other models of supersymmetry breaking. For example, the parameters $A$ and $B$ vanish at a high scale in the gauge mediated models \cite{Giudice:1998bp} with the minimal messenger sector and get  generated purely radiatively. In such scenarios, $\frac{{\rm Det} M_R}{m_{3/2}^3}$ can be naturally small offering the possibility of raising the value of $m_{3/2}$ compared to the gaugino masses.

One can also use the fitted value of $m$  and $\xi$ to determine the Yukawa coupling $\lambda_\tau^2$. Using Eq. (\ref{m}) and neglecting small $\zeta$ dependent term, one finds 
\beqa \label{lamtau}
\lambda_\tau^2 &\simeq & \frac{m \xi \lambda^2}{A_0 k^2 C_e^2 m_{3/2}^2} = \frac{\mu^2}{\mu_0^2} \frac{\lambda^2}{C_e^2}\,\frac{m \xi}{A_0 v_2^2}\,,\eeqa
where the second equality is obtained using the value of $k$ from Eq. (\ref{ktau2}). It is to be noted that $\lambda_\tau^2$ as given above is determined by the $R$ parity violating contribution $m\xi$ rather than by the seesaw contribution $m$. For $\mu \sim \mu_0$, $A_0 = (2 M_g)^{-1}$ and $C_e = \lambda = 1$, we get
\beqa \label{lamtau_num}
\lambda_\tau^2 &\sim & 2 \times 10^{-11}\, \left(\frac{m \xi}{0.1\, {\rm eV}} \right)\, \left(\frac{100\,{\rm GeV}}{v_2} \right)^2\,\left(\frac{M_g}{1\, {\rm TeV}} \right)\,.\eeqa
This is to be compared with the expectation based on the seesaw, $\lambda_\tau^2\sim \frac{m_\nu M_R}{v^2}\sim 10^{-11} $, for a TeV scale $M_R$ and $m_\nu \sim 0.1$ eV. The similarity of the  magnitudes of $\lambda_\tau^2$ obtained through these two different approaches is not a coincidence since the model requires both these contributions to be of similar order for the realistic neutrino mass spectrum.

\begin{figure}[t]
\centering
\subfigure{\includegraphics[width=0.42\textwidth]{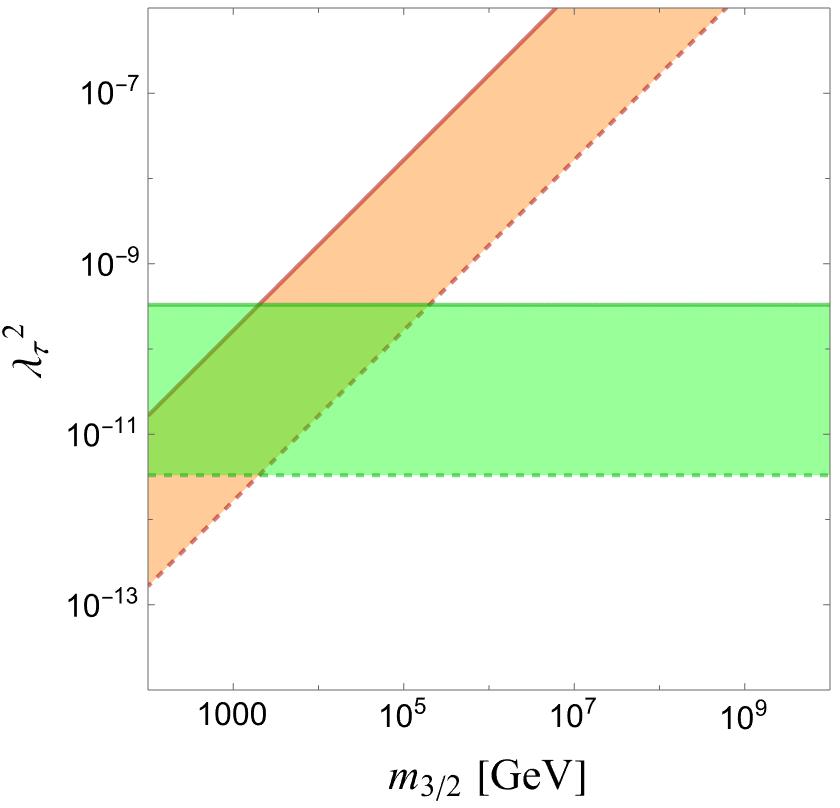}}\hspace*{0.6cm}
\subfigure{\includegraphics[width=0.42\textwidth]{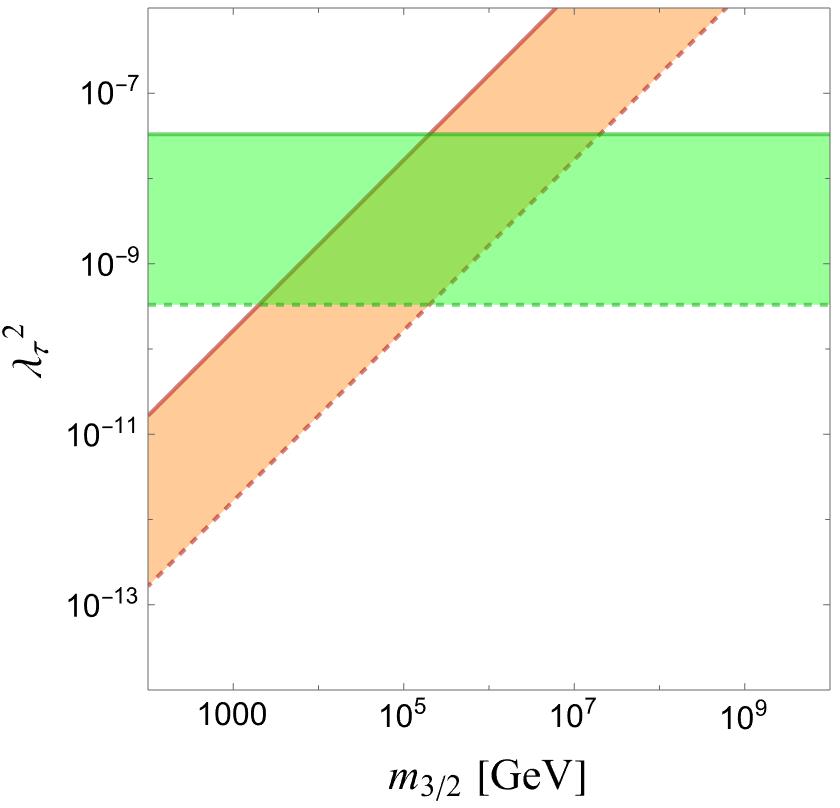}}
\caption{Contours of $m = 0.01$ eV (red dashed), $m = 1$ eV (red solid), $m \xi = 0.01$ eV (green dashed), $m = 1$ eV (green solid) for various values of $m_{3/2}$ and $\lambda_\tau^2$ and for gaugino mass $M_g =10^3$ GeV (left panel) and $M_g = 10^5$ GeV (right panel). Only in the overlapping region, a realistic light neutrino mass spectrum can be reproduced.} 
\label{fig1}
\end{figure}

The above observations can also be summarized by comparing the overall scales of the seesaw and $R$-parity contributions to the light neutrino masses. The first is proportional to the parameter $m$ given in Eq. (\ref{m}) which goes as $\lambda_\tau^2 v_2^2/m_{3/2}$. The second is determined as $m \xi$ and it is independent of $m_{3/2}$ for the fixed gaugino mass scale $M_g$ as can be observed from the expression of $\xi$ in Eq. (\ref{xi_simpl_2}). The requirement that $m$ and $m \xi$ are of a similar order, therefore, puts an upper bound on $m_{3/2}$ as it is shown in Fig. \ref{fig1}. Further, if one chooses $M_g \simeq m_{3/2}$, both the contributions scale as $\lambda_\tau^2 v_2^2/m_{3/2}$ leading to a situation similar to the one obtained in the case of conformal $W$ discussed in section \ref{sec:model}. 

\section{Majoron-neutrino couplings}
\label{sec:majoron_neutrino}
In the present framework, there exists a Goldstone boson of the spontaneously broken LN symmetry, namely the Majoron \cite{Chikashige:1980ui}. It can be identified as
\be \label{J}
J \simeq \frac{1}{\sqrt{2}} \left(\tilde{N}_{I \mu} - \tilde{N}_{I e} \right) + \frac{\omega^\prime_\alpha}{U}\, \tilde{\nu}_{I \alpha}\,,\ee
where $U = \sqrt{\langle \tilde{N_e} \rangle^2 + \langle \tilde{N_\mu} \rangle^2}$ is the scale of the $U(1)_L$ breaking and the subscript $I$ denotes the imaginary part of the corresponding scalar field. The Majoron is an $SU(2)$ singlet and does not directly couple to the SM neutrinos. However, this coupling is generated through (a) light-heavy neutrino mixing and (b) neutrino-neutralino mixing. This coupling can be parametrized as
\be \label{nuJ}
{\cal L}_{\nu J} = - \frac{i}{2} g_{ij}\, \nu^T_{L i} C^{-1} \nu_{L j}\, J\, + {\rm h.c.}\,,\ee
in the physical neutrino basis.

The strongest limits on $g_{ij}$ arise from the precision measurement of the acoustic peak of the CMB spectrum which requires that neutrinos must be free streaming at the time of photon decoupling and therefore cannot couple strongly to a nearly massless Majoron. This requirement translates into \cite{Hannestad:2005ex,Raffelt:1994ry,Ayala:2014pea}
\beqa \label{gijlimit}
|g_{ii}| & \lesssim &10^{-7}\,,\nonumber \\
|g_{ij}| & \lesssim & 0.61 \times 10^{-11}\, \left(\frac{0.05\, {\rm eV}}{m_{\nu_j}}\right)^2\,,\eeqa
for $i \neq j$ and $m_{\nu_j}>m_{\nu_i}$. We now show that these limits are satisfied in the model in spite of relatively low LN breaking scale $\sim m_{3/2}$ without conflicting with constraints from neutrino masses and mixing.

Contribution to $g_{ij}$ arising from the light-heavy neutrino mixing in the present model can be estimated as follows. Eq. (\ref{W}) gives rise to the following interaction between the heavy neutrinos and the Majoron:
\be \label{nuJa1}
{\cal L}^{(a)}_{\nu J} =- \frac{i \lambda}{\sqrt{2}}\, \left( N_e^T C^{-1} N_\tau\, \tilde{N_\mu}_{I}  + N^T_\mu C^{-1} N_\tau\, \tilde{N_e}_{I}\right) + {\rm h.c.}\,. \ee
Subsequently, the couplings with light neutrinos can be estimated using the diagonalization of the relevant blocks of ${\cal M}_{\cal N}$. The latter can be block diagonalized at the leading order by a unitary matrix,
\be \label{Uneut}
U_{\cal N} = \left( \ba{ccc} 1  & \rho_{\nu N} & \rho_{\nu \chi}\\ 
								-\rho_{\nu N}^\dagger & 1 & \rho_{N \chi}\\
								-\rho_{\nu \chi}^\dagger & -\rho_{N \chi}^\dagger & 1 \ea \right)\,+\, {\cal O}(\rho^2)\,,
\ee
such that $U_{\cal N}^T {\cal M}_{\cal N} U_{\cal N}$ is a block diagonal matrix. Here,
\be \label{rhos}
\rho_{\nu N} \simeq (m_{\nu N} M_R^{-1})^*\,,~~
\rho_{\nu \chi} \simeq (m_{\nu \chi} M_\chi^{-1})^*\,,~~
\rho_{N \chi} \simeq (m_{N \chi} M_\chi^{-1})^*\,.\ee

The neutral fermions in the block-diagonal basis are given by $U_{\cal N}^\dagger \psi^0$. Writing the light neutrinos in block diagonal basis as $\nu_\alpha^\prime$ and using Eq. (\ref{rhos}), one finds
\be \label{Nnu_mixing}
N_e = -(\rho_{\nu N}^\dagger)_{1 \alpha} \, \nu_{L \alpha}^\prime\,,~~N_\mu = -(\rho_{\nu N}^\dagger)_{2 \alpha}\, \nu_{L \alpha}^\prime\,,~~N_\tau = -(\rho_{\nu N}^\dagger)_{3 \alpha}\, \nu_{L \alpha}^\prime\,. \ee
Substituting the above in Eq. (\ref{nuJa1}) and using the definition of $J$ in Eq. (\ref{J}), we find the light neutrino-Majoron interaction as
\be \label{nuJa2}
{\cal L}^{(a)}_{\nu J} = -\frac{i}{2} g_{ij}^{(a)}\, \nu^T_{L i} C^{-1} \nu_{L j}\, J\, + {\rm h.c.}\,,\ee
with
\beqa \label{ga}
g^{(a)}_{ij} &=& \lambda \left((\rho_{\nu N}^\dagger)_{1 \alpha} - (\rho_{\nu N}^\dagger)_{2 \alpha}\right) (\rho_{\nu N}^\dagger)_{3 \beta}\,U_{\alpha i} U_{\beta j}\, \nonumber \\
&=& \frac{\lambda C_e v_2^2}{C_\tau m_{3/2}^2 (C_\tau \tilde{C}_\tau - 2 C_e^2)}\,\lambda_\alpha \lambda_\beta\,U_{\alpha i} U_{\beta j}\,\nonumber \\
&=& -\frac{\lambda \tilde{m}_0}{C_e m_{3/2}}\,\lambda_\alpha \lambda_\beta\,U_{\alpha i} U_{\beta j}\,,\eeqa
where we use Eq. (\ref{tildem0}) to arrive at the last equality. In the above, we have used $\nu_{L \alpha}^\prime = U_{\alpha i} \nu_{Li}$ where $\nu_{Li}$ are the mass eigenstates which are identified as physical light neutrinos. The couplings $g^{(a)}_{ij} $ are suppressed by $\tilde{m}_0/m_{3/2}$ as a consequence of heavy-light mixing.

The second contribution arises from the Yukawa interaction in Eq. (\ref{W}) and through gauge interactions quantified by the following terms:
\beqa \label{nuJb1}
{\cal L}^{(b)}_{\nu J} &=& - \frac{i}{\sqrt{2}} \lambda_\alpha\, \nu_{L \alpha}^{\prime T} \tilde{N_e}_{I} C^{-1} \tilde{h}_2^0 - \frac{i}{2} (g^\prime \tilde{B}^T - g \tilde{W}^{0T}) C^{-1} \nu^\prime_{L \beta}\, \tilde{\nu}_{I \beta}+{\rm h.c.}\,, \eeqa
where $-i\lambda_1 \equiv \tilde{B}$ and $-i\lambda_2 \equiv \tilde{W}^0$. The Higgsino, Bino and Wino contain the light neutrinos through mixing. From the block diagonalization, one finds
\beqa \label{chinu_mixing}
\tilde{h}_2^0 &=& -(\rho_{\nu \chi}^\dagger)_{4 \alpha} \, \nu_{L \alpha}^\prime = \frac{v_1}{\mu} A_0\, \omega_\alpha \, U_{\alpha i}\, \nu_{L i}\,, \nonumber \\
\tilde{B} &=& -(\rho_{\nu \chi}^\dagger)_{1 \alpha}\,\nu_{L \alpha}^\prime = - \frac{\sqrt{2} g^\prime M_2}{g^2 M_1 + g^{\prime 2} M_2}\,A_0\, \omega_\alpha\, U_{\alpha i}\, \nu_{L i} \,, \nonumber \\
\tilde{W}^0 &=& -(\rho_{\nu \chi}^\dagger)_{2 \alpha}\, \nu_{L \alpha}^\prime = \frac{\sqrt{2} g M_1}{g^2 M_1 + g^{\prime 2} M_2}\,A_0\, \omega_\alpha\, U_{\alpha i}\, \nu_{L i} \,, \eeqa
Substituting the above in Eq. (\ref{nuJb1}) and using Eq. (\ref{J}), we find
\be \label{nuJb2}
{\cal L}^{(b)}_{\nu J} = -\frac{i}{2} g_{ij}^{(b)}\, \nu^T_{L i} C^{-1} \nu_{L j}\, J\, + {\rm h.c.}\,,\ee
where
\be \label{gb}
g^{(b)}_{ij} =  - \frac{A_0 \lambda}{C_e m_{3/2}} \omega_\alpha \omega_\beta\,U_{\alpha i} U_{\beta j}\,,\ee 
where $A_0$ is as defined in Eq. (\ref{A0}).

Combining the two contributions and using the definition Eq. (\ref{nuJ}), the $3\times 3$ light neutrino-Majoron coupling matrix can be written as
\be \label{gnuJ}
g = -\frac{\lambda }{C_e m_{3/2}}\,\, U^T\, \left(\tilde{m}_0{\cal A} + A_0 {\cal B}\right)\,U \,.\ee
Note that the quantity in the bracket above is the effective neutrino mass matrix $m_\nu$ in the limit $\eta \to 0$. Therefore, using the simplified expression of $m_\nu$ obtained in Eq. (\ref{mnu_ana_simp}) with $\eta = 0$, we can express
\be \label{gnuJ2}
g =g_0\,\,  U^T\,\left(
\begin{array}{ccc}
 r_e^2 \left(1+\xi_0  s_e^2 \right) & r_e r_{\mu } \left(1+\xi_0 s_e s_{\mu }\right) & r_e \left(1+\xi_0 s_e\right) \\
 r_e r_{\mu } \left(1+\xi_0 s_e s_{\mu }\right) & r_{\mu }^2 \left(1+\xi_0 s_\mu^2\right) & r_{\mu } \left(1+ \xi_0 s_{\mu }\right) \\
r_e \left(1+\xi_0 s_e\right)  & r_{\mu } \left(1+ \xi_0 s_{\mu }\right) & 1+ \xi_0 \\
\end{array}
\right)\, U \,,\ee
with
\be \label{g0}
g_0 = -\frac{\tilde{m}_0 \lambda_\tau^2}{\langle \tilde{N}_e \rangle}\,, \ee
and $\xi_0$ as defined in Eq. (\ref{ratio}). When $\eta=0$, the matrix in the bracket in Eq. (\ref{gnuJ2}) is identified with $m_\nu$ and $g$ becomes a diagonal matrix leading to only $g_{ii}$ nonzero. This case is relatively poorly constrained as seen from the limits in Eq. (\ref{gijlimit}).

Numerically, all the dimensionless parameters appearing in the matrix shown in Eq. (\ref{gnuJ2}) are of order one as can be seen from the numbers in Tables \ref{tab:sol_NH} and \ref{tab:sol_IH}. The overall scale of neutrino-majoron coupling, therefore, is given by $g_0$. Also noting that the numerator in Eq. (\ref{g0}) is $m$ in the limit $\eta \to 0$, one can estimate an overall size of $g_0$ as
\beqa \label{g0m}
|g_0| & \simeq &  \frac{m}{|\langle \tilde{N}_e \rangle|} = 10^{-11}\, \left(\frac{m}{1\,{\rm eV}}\right) \left(\frac{100\,{\rm GeV}}{|\langle \tilde{N}_e \rangle|}\right)\,.\eeqa
It is seen that $|g_0|$ remains suppressed for a large range in the parameter space. The matrix appearing in Eq. (\ref{gnuJ}) is also completely determined by the fits of the table since $\xi_0 \approx \xi$. For example, for solution NH8 listed in Table \ref{tab:sol_NH}, we get 
\be 
|g| \simeq 10^{-13}\, \left(\frac{100\,{\rm GeV}}{|\langle \tilde{N}_e \rangle|}\right)\, \left(
\begin{array}{ccc}
 0 & 0 & 0 \\
 0 & 1.79624 & 2.55013 \\
 0 & 2.55013 & 1.52902 \\
\end{array}
\right)\,.\ee
We find that all the solutions listed in Tables \ref{tab:sol_NH} and \ref{tab:sol_IH} satisfy the constraints, Eq. (\ref{gijlimit}), for $|\langle \tilde{N}_e \rangle| \gtrsim 100$ GeV.

\section{Majoron-Charged lepton couplings}
\label{sec:majoron_cl}
As in the case of the neutrinos, the charged leptons $e_\alpha$ do not directly couple to Majoron but these couplings arise from their mixing with the charginos. In general, the Majoron-charged lepton couplings can be parametrized as
\be \label{h_def}
{\cal L}_{ej} = i\,h_{\alpha \beta}\,e_\alpha e^c_\beta J\,+\,{\rm h.c.}\,.\ee
Majoron has both the diagonal and the off-diagonal couplings. The latter is constrained from the flavour violating decays such as $e_\alpha \rightarrow e_\beta\,J$ \cite{Albrecht:1995ht,PhysRevD.34.1967}. Non-observation of these implies \cite{Raffelt:1994ry,Ayala:2014pea}:
\be\label{con1}
|h_{e\mu}|<1.9\times 10^{-11}\,,~~|h_{\tau e}|,|h_{\tau \mu}|<1.1\times 10^{-7}\,.\ee
The requirement that the red giant stars do not lose energy through Majoron emission from electrons   constrain $|h_{ee}|$, namely
\be \label{con2}
|h_{ee}|\leq 2.57\times 10^{-13}\,.\ee
Despite such strong constraints, a Majoron with $U(1)_L$ breaking scale as low as $\sim 100$ GeV can satisfy the above limits as we discuss in detail below.

The three charged lepton and two chargino fields can be represented as
\be 
\psi_-=\left(\ba{c}
e_\alpha\\
\chi_m\ea \right)\,,~~~\psi_+=\left(\ba{c}
e_\alpha^ c\\
\chi_m^ c\ea \right)\ee
with
\be \chi_m=\left(\ba{c}
-i \lambda^-\\
h_1^-\ea \right)\,,~~~\chi_m^c=\left(\ba{c}
-i \lambda^+\\
h_2^{+}\ea \right)\ee
The charged fermion mass term is parametrized by
\be\label{charginomass}
-{\cal L}_c = \psi_-^{ T\,}{\cal M}_c\, \psi_+\,+\,{\rm h.c.}\,, \ee
where $5\times 5$ matrix ${\cal M}_c$ is defined as
\be \label{Mc}
{\cal M}_c=\left(\ba{cc}
m_l&m_{lc}\\
m_{lc}^\prime&M_c\\
\ea \right)~.\ee
Here, $m_l$ denotes the (diagonal) charged lepton mass matrix before mixing with charginos.
$m_{lc}$, $m_{lc}^\prime$ arise from the  $R$-parity violation and are given by
\be \label{mlmlc}
m_{lc}=\left(\ba{cc}
g \omega_e^\prime&\epsilon_e\\
g \omega_\mu^\prime&\epsilon_\mu\\
g \omega_\tau^\prime&\epsilon_\tau\\ \ea \right)\,,~~
m_{lc}^\prime=\left( \ba{ccc}
0&0&0\\
\lambda_e^\prime \omega_e^\prime&\lambda_\mu^\prime \omega_\mu^\prime&\lambda_\tau^\prime \omega_\tau^\prime\\
\ea\right)\,.
\ee
$M_c$ is the usual chargino mass matrix in the MSSM,
\be \label{MSSMMC}
M_c=\left(\ba{cc}M_2&g v_1\\
g v_2&\mu\\ \ea \right)~.\ee

Next, we define a basis
\be \label{psiprime}
\psi_\pm^\prime = U_\pm^\dagger\, \psi_\pm\,,\ee
such that ${\cal M}_c$ assumes a block diagonal form in this basis. Explicitly,
\be \label{blockd}
U_-^T{\cal M}_cU_+=\left(\ba{cc}
m^{\rm eff}_l&0_{3\times 2}\\
0_{2\times 3}&M^{\rm eff}_c \\ \ea\right)\,.\ee
The $U_\pm$ can be parameterized as
\be \label{upm}
U_\pm=\left(\ba{cc}
\sqrt{1-B_\pm B_\pm^\dagger} &B_\pm\\
-B_\pm^\dagger&\sqrt{1-B_\pm^\dagger B_\pm }\\
\ea \right)\,.\ee
It is always possible \cite{Grimus:2000vj} to choose $3\times 2$ matrices $B_\pm$ which puts ${\cal M}_c$ in the block diagonal form of Eq. (\ref{blockd}).

The coupling between the Majoron and charged lepton can be directly obtained from the basic Lagrangian as done in the case of neutrinos. Alternatively, one can use the effective Lagrangian describing the coupling of  Majoron to the divergence of the 
broken $U(1)$ current $j^\mu$,
\be \label{eeJ1}
-{\cal L}_{eJ}=-\frac{i}{U}\, J\, \partial_\mu j^\mu~.\ee
Here, 
\be \label{jmu}
j^\mu=-\bar{e}_\alpha\bar{\sigma}^\mu e_\alpha+\bar{e}^c_\alpha\bar{\sigma}^\mu e^c_\alpha~\ee
is the leptonic current in the flavour basis.  $U$ denotes the $U(1)_L$ symmetry breaking scale as already defined in the previous section. The mixing terms $m_{lc}$ and $m_{lc}^\prime$
violate $U(1)_L$ and hence contribute to $\partial_\mu j^\mu$. This contribution is given by
\be \label{div1}
\partial_\mu j^\mu=i\, \psi^T_-\left(q{\cal M}_c+{\cal M}_c q^c\right)\psi_+\, +\,{\rm h.c.}\,,\ee
where
\be \label{qpm}
q=\left(\ba{cc}
{1}_{3\times 3}&{ 0}_{3\times 2}\\
{0}_{2\times 3}&{0}_{2\times 2}\ea \right)~~~,~~~q^c=\left(\ba{cc}
-{1}_{3\times 3}&{0}_{3\times 2}\\
{0}_{2\times 3}&{0}_{2\times 2}\ea \right)\, \ee
are the diagonal charge matrices representing $U(1)_L$ charges of the underlying charged leptons and charginos.
Eq. (\ref{div1}) can be rewritten in the block diagonal basis as
\be \label{div2}
\partial_\mu j^\mu=i\psi^{\prime T}_-U_-^T\left(q {\cal M}_c+q^c {\cal M}_c\right)U_+\psi_+^\prime\, +\,{\rm h.c.}\,.\ee
Using Eqs. (\ref{blockd},\ref{upm}), the Majoron charged lepton coupling can be written in the block diagonal basis as
\beqa \label{eeJ2}
{\cal L}_{eJ} &=&\frac{i}{\sqrt{2}\vev{\tilde{N}_e}}\left(m^{\rm eff}_l\, B_+B_+^\dagger-B_-^*B_-^T\,m^{\rm eff}_l\right)_{\alpha\beta}\,e_\alpha e_\beta^c J\, +\,{\rm h.c.}\,.\eeqa
Comparing with the definition Eq. (\ref{h_def}), we get
\be \label{h_expression}
h_{\alpha \beta} = \frac{1}{\sqrt{2}\vev{\tilde{N}_e}}\left(m^{\rm eff}_l\, B_+B_+^\dagger-B_-^*B_-^T\,m^{\rm eff}_l\right)_{\alpha\beta}\,.\ee
This equation is valid to all orders in the seesaw expansion. $m^{\rm eff}_l$ is the effective charged lepton mass matrix whose eigenvalues correspond to the masses of the charged leptons.

$B_{\pm}$  can be iteratively determined using Eq. (\ref{blockd}) in the seesaw approximation $m_l < m_{lc},m_{lc}^\prime,\ll M_C$. At the leading order, one finds \cite{Grimus:2000vj}
\be \label{first} 
B_+\approx(M_c^{-1}m_{lc}^\prime)^\dagger\,,~~B_-\approx (m_{lc}M_c^{-1})^*\,,~~m_l^{\rm eff} \approx m_l-m_{lc}M_c^{-1}m_{lc}^\prime\,.\ee
Substitution of the above in Eq. (\ref{h_expression}) indicates that the charged lepton-Majoron couplings, at the leading order, are suppressed by two powers of seesaw expansion parameter $m_{l c} M_c^{-1}$ or $m^\prime_{lc} M_c^{-1}$. This is different from neutrino-Majoron coupling which goes as $\sim m_{\nu N} M_R^{-1}$ at the leading order. The model, therefore, predicts additionally suppressed coupling between the Majoron and the charged leptons.

Explicit computation of $h_{\alpha \beta}$ using the leading order expressions of $B_{\pm}$ and neglecting the terms of ${\cal O}(v_{1,2}/\mu)$, we get
\beqa \label{h_leading}
h_{\alpha \beta} = - \frac{1}{\sqrt{2}\vev{\tilde{N}_e}}\,\left(m_\beta \left(\frac{\epsilon _\alpha \epsilon _\beta}{\mu^2} + g^2 \frac{\omega^\prime _\alpha \omega^\prime _\beta}{M_2^2}  \right) - m_\alpha \frac{\lambda _\alpha^\prime \lambda _\beta^\prime \omega^\prime _\alpha \omega^\prime _\beta}{\mu ^2}  \right)\,.\eeqa
Using Eq. (\ref{basis}) and noting that $k_\alpha$ of Eqs. (\ref{kalpha1},\ref{kalpha2}) are smaller than one, the dominant contribution to $h_{\alpha\beta}$ comes from the term proportional to $\epsilon_\alpha\epsilon_\beta$. This gives, for example,
\beqa 
h_{ee} &\simeq & r_e^2\lambda_\tau^2m_e\frac{\vev{\tilde{N}_e}}{\mu^2}~,\nonumber\\
& = & 4  \times 10^{-16}\,\left(\frac{\lambda_\tau^2}{10^{-10}}\right)\left(\frac{\vev{\tilde{N}_e}}{100 \,{\rm GeV}}\right)\left(\frac{500\,{\rm GeV}}{\mu}\right)^2\,,\eeqa
where we take $r_e\equiv\frac{\lambda_e}{\lambda_\tau}=4.44$ which is the largest value among all the solutions listed in Tables \ref{tab:sol_NH}  and \ref{tab:sol_IH}. The obtained value of $h_{ee}$
is far below the  constraint (\ref{con2}). Similarly, the off-diagonal elements are also found much below the existing constraint, Eq. (\ref{con1}).

\section{Right-handed neutrinos and direct search prospects}
\label{sec:RHN_spect}
The three RH neutrinos in the model have masses close to the electroweak scale and they can be produced at colliders if their mixing with the other particles is favourable. In this section, we discuss such possibilities in detail and point out ranges of parameters of the model where the direct detection of RH neutrinos is possible. There are two standard methods of detecting the RH neutrinos, see \cite{Deppisch:2015qwa} for a review. The RH neutrinos can couple to the standard $W$ and $Z$ through their mixing with the active neutrinos. The same mixing also causes the decay of the $N_i$ into charged leptons. This leads to dilepton + 2 jet signals  which are free from the SM background. Such signals however depend on the fourth power of the RH mixing $|V_{\alpha N_i}|$ and are therefore sensitive to relatively large mixing only. It is also possible to probe slightly smaller $|V_{\alpha N_i}|$ through a technique involving displaced vertices at the LHC, see for example \cite{Helo:2013esa,Izaguirre:2015pga,Gago:2015vma,Antusch:2017hhu,Cottin:2018nms,Abada:2018sfh,Drewes:2019fou,Liu:2019ayx,Drewes:2019vjy,Das:2019fee}.

In the $R$-parity violating  supersymmetric models, the RH neutrinos can also mix with the neutralinos and this mixing could be much larger than $|V_{\alpha N_i}|$. In such a situation, the RH neutrinos are produced in the decays of neutralinos and further decay through their mixing with the active neutrinos. Relatively small mixing with active neutrinos can result in the delayed decays of the RH neutrinos which lead to the distinctive displaced vertex signatures as discussed in \cite{Lavignac:2020yld}. We focus on this possibility and discuss it at length in this section.

The RH neutrino mass eigenstates $N_i$ are related to the flavour states $N_\alpha$ as
\be \label{ninalpha}
N_\alpha =(U_R)_{\alpha i}N_i\,,\ee
where $U_R$ diagonalizes the mas matrix $M_R$ given in Eq. (\ref{MR}). Explicitly, $U_R$ is given by
\be \label{UR}
U_R \simeq \left(
\begin{array}{ccc}
 \frac{i}{\sqrt{2}} & \frac{\cos \theta}{\sqrt{2}} & \frac{\sin \theta}{\sqrt{2}} \\
 -\frac{i}{\sqrt{2}} & \frac{\cos \theta}{\sqrt{2}} & \frac{\sin \theta}{\sqrt{2}} \\
 0 & -\sin \theta &  \cos \theta \\
\end{array}
\right)\,,\ee
Mixing of $N_i$ with the charged leptons follows from Eqs. (\ref{Uneut},\ref{ninalpha}). Specifically, one can write
\be\label {Vdef}
\nu'_\alpha=(\rho_{\nu D})_{\alpha\beta}N_\beta=(\rho_{\nu D})_{\alpha\beta}(U_R)_{\alpha i}N_i\equiv V_{\alpha N_i }N_i\,.\ee
Using the above relation, we find
\beqa \label{valphan}
V_{\alpha N_1}&=&=\frac{-i\lambda _\alpha v_2}{\sqrt{2} M_{N_1}}=\frac{-ir_\alpha \lambda _\tau v_2}{\sqrt{2} M_{N_1}}~,\nonumber\\
V_{\alpha N_2}&=&\frac{c_\theta\lambda_\alpha v_2}{\sqrt{2} M_{N_2}}=\frac{c_\theta r_\alpha \lambda_\tau v_2}{\sqrt{2} M_{N_2}}~,\nonumber\\
V_{\alpha N_3}&=&\frac{s_\theta\lambda_\alpha v_2}{\sqrt{2} M_{N_3}}=\frac{s_\theta r_\alpha \lambda_\tau v_2}{\sqrt{2} M_{N_3}}\,.\eeqa

The predicted RH mixing, $V_{\alpha N_i}\sim \frac{(m_{\nu N} U_R)_{\alpha i}}{M_{N_i}}$ is in agreement with a typical seesaw scenario. For $\lambda_\tau \sim 10^{-5}$ and $M_{N_i}\sim 100$ GeV, one gets $V_{\alpha N_i}\sim r_\alpha\, 10^{-7}$. This is much smaller than the current limit from the direct searches, $|V_{\alpha N}|^2 < 10^{-5}$ \cite{ATLAS:2019kpx,CMS:2018iaf}, see for example, the left panel in Fig. \ref{fig3}. One can raise $V_{\alpha N_i}$ for a smaller $M_{N_i}$ and a larger $\lambda_\tau$. For example, $\mu/\mu_0 \sim 5$ and $M_{N_1}=100 $ GeV in Eq. (\ref{lamtau}) gives $\lambda_\tau\sim 2 \times 10^{-3}$ and leads to $V_{e N_3}\sim 3 \times 10^{-3} \sin\theta$. While this is much larger than the typical TeV scale seesaw prediction, this still falls short of the existing limits from the collider searches. Nevertheless, it is possible to probe such or even smaller mixing through the specific displaced vertex signature as suggested in \cite{Lavignac:2020yld} provided that RH neutrinos are produced through neutralinos.

The mixing between heavy neutrinos $N_i $ and neutralinos, relevant for displaced vertex signatures at the colliders, can be estimated using
\be 
N_\alpha = \left(\rho_{N\chi}\right)_{\alpha m}\,(\psi_0)_m\,.\ee
Using Eqs. (\ref{rhos},\ref{Mneut}) and ignoring the term $\lambda_\alpha \omega^\prime_\alpha$ in $m_{N \chi}$, we find that only the state $N_\tau$ has mixing with Higgsinos and gauginos at the leading order. It is given by
\beqa \label{Xchi_mix}
N_\tau& \simeq & \frac{2 \eta A_0 v M_1 M_2}{\mu (g^2 M_1 + g^{\prime 2} M_2)} \left(\cos\beta\, \tilde{h}^0_1 + \sin\beta\, \tilde{h}^0_2 + \frac{g^\prime v}{\sqrt{2} M_1}\cos 2\beta\, \tilde{B} -\frac{g v}{\sqrt{2} M_2}\cos 2\beta\, \tilde{W^0}  \right). \eeqa
Overall, the couplings with Higgsinos are suppressed by $v/\mu$ only while the ones with Bino (Wino) are suppressed by additional power of $v/M_1$ ($v/M_2$). Replacing $N_\tau$ by physical states through Eq. (\ref{UR}), we find the heavy neutrino-Higgsino mixing as
\be \label{VNh}
V_{N_2 \tilde{h}_1^0} \simeq - \frac{2 \eta A_0 v M_1 M_2}{\mu (g^2 M_1 + g^{\prime 2} M_2)}\,\cos\beta \sin\theta\,,~~~V_{N_3 \tilde{h}_1^0} \simeq \frac{2 \eta A_0 v M_1 M_2}{\mu (g^2 M_1 + g^{\prime 2} M_2)}\,\cos\beta \cos\theta\,.\ee
As it can be seen, $N_1$ does not mix with Higgssinos at the leading order.

\begin{figure}[t]
\centering
\subfigure{\includegraphics[width=0.42\textwidth]{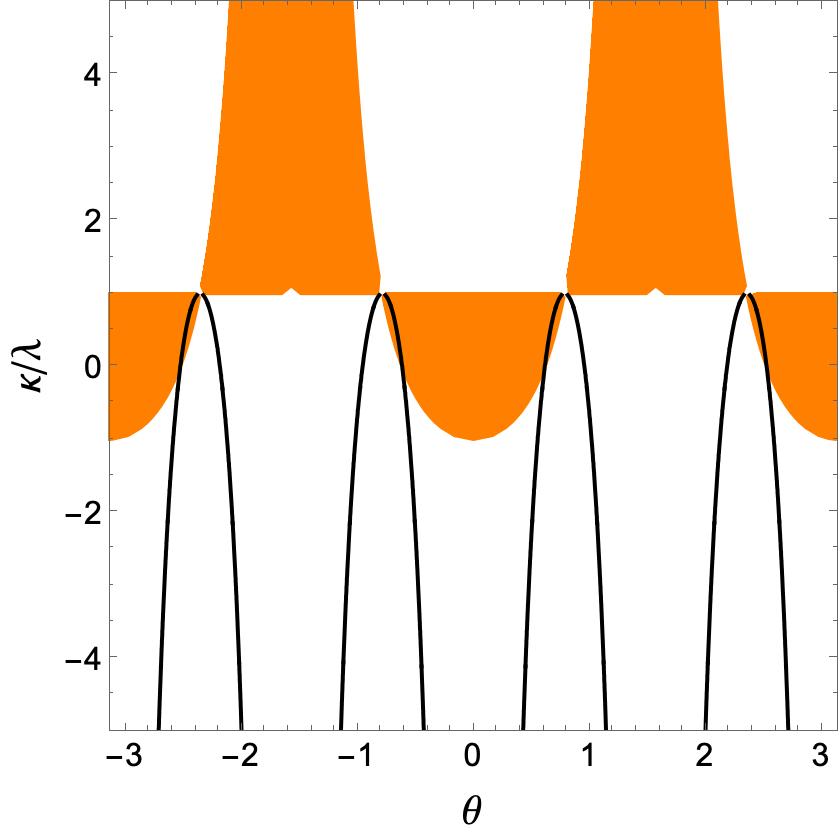}}
\caption{Regions in $\theta$ and $\kappa/\lambda$ leading to $M_{N_3} < M_{N_{1,2}}$. The black contours indicate the correlation, Eq. (\ref{sp}), resulting from a specific choice of soft parameters.} 
\label{fig2}
\end{figure}

\begin{figure}[t]
\centering
\subfigure{\includegraphics[width=0.42\textwidth]{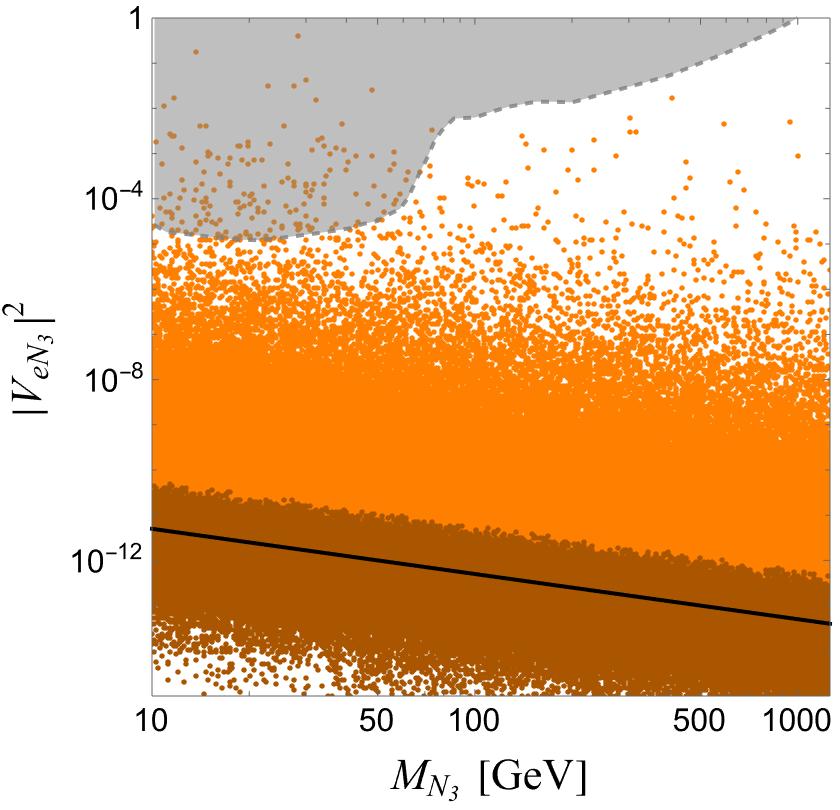}}\hspace*{0.6cm}
\subfigure{\includegraphics[width=0.42\textwidth]{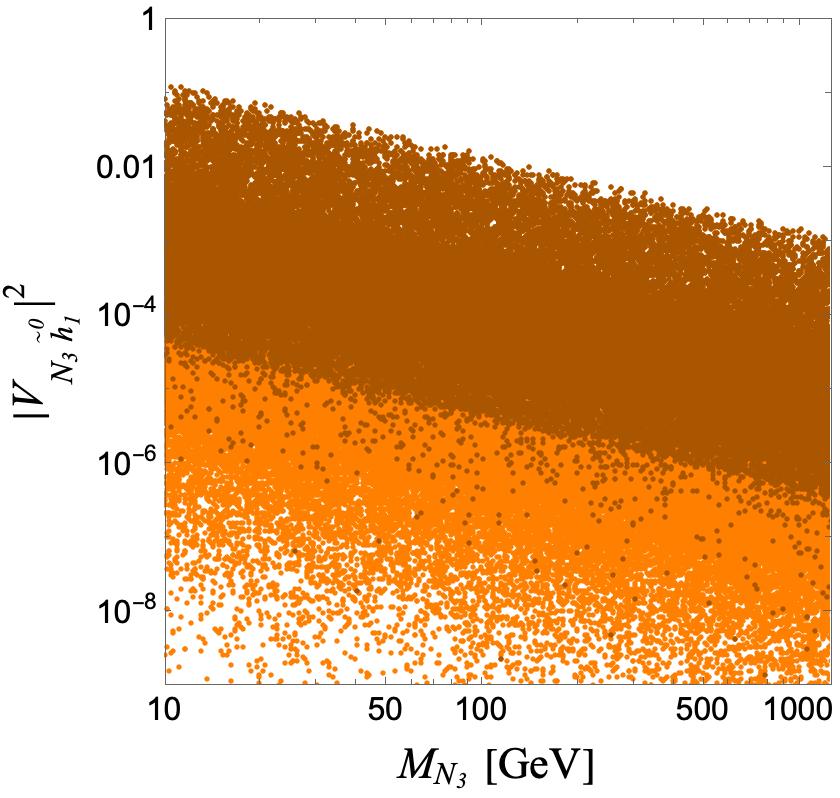}}
\caption{Variations of $|V_{eN_3}|^2$  ($|V_{N_3 \tilde{h}_1^0}|^2$) with $M_{N_3}$ in the left (right) panel, for example, values of various parameters as described in the text. The darker points correspond to cases in which  $|V_{N_3 \tilde{h}_1^0}|^2/|V_{eN_3}|^2 \geq 10^8$. The solid black line in the left panel depicts the values of $|V_{\alpha N}|^2$ expected from a typical seesaw relation, $|V_{\alpha N}|^2=\frac{0.05\, {\rm eV}}{M_{N}}$. The grey region covered by dashed contour is excluded at $95 \%$ C.L. by the direct searches of heavy neutral leptons \cite{ATLAS:2019kpx,CMS:2018iaf}.} 
\label{fig3}
\end{figure}

The states $N_2$ and $N_3$ can be efficiently produced in the decays of Higgsinos provided that $M_{N_{2,3}}$ are smaller than the Higgsino masses. Motivated by this, we look for the values of parameters for which $N_3$ is the lightest among all three RH neutrinos\footnote{The role of $N_2$ and $N_3$ can be interchanged by changing the sign of $\cos 2\theta$, see Eq. (\ref{RH_masses}).}. As can be seen from Eq. (\ref{RH_masses}), the ratios $M_{N_3}/M_{N_1}$ and $M_{N_3}/M_{N_2}$ depend only on $\theta$ and $\kappa/\lambda$. By varying the values of these parameters in reasonable ranges, we look for the region in which $M_{N_3}$ is the lightest RH neutrino. The results are displayed in Fig. \ref{fig2} in which the regions shaded in orange correspond to $M_{N_3} < M_{N_{1,2}}$.
In the same figure, we also show the contours corresponding to 
\be \label{sp}
\tan^2 2\theta = \frac{8}{1-\frac{\kappa}{\lambda}}\,.\ee
The above arises from a relation between $C_e$ and $C_\tau$ obtained as Eq. (\ref{Ce_Ctau}) as a result of a specific choice of the soft parameters.

Next, for $M_{N_3} < M_{N_{1,2}}$, we estimate the values of mixing parameters $V_{e N_3}$ and $V_{N_3 \tilde{h}_1^0}$ given in Eq (\ref{valphan}) and (\ref{VNh}), respectively. This requires the specification of several parameters. Using Eqs. (\ref{newmu},\ref{RH_masses}), we set
\be \label{eta_RH}
\eta = \lambda\, \frac{\mu_0 - \mu}{M_{N_1}} = \lambda\, \frac{\mu_0 - \mu}{ 2M_{N_3}}\,\left(1+\frac{\kappa}{\lambda} + \left(\frac{\kappa}{\lambda} - 1 \right) \sec 2\theta \right)\,.\ee
For $\lambda_\tau$, we use the expression obtained in Eq. (\ref{lamtau}) and replace $C_e$ using Eq. (\ref{t2th}). Also,  $\lambda$ can be determined from Eq. (\ref{xi_simpl}) and using the leading order expression of $\xi$ and also using Eqs. (\ref{RH_masses},\ref{detMR}). It leads to 
\be \label{lambda_result}
\lambda^2 \simeq 2 M_{N_{3}}\frac{A_0}{\xi}\left(\frac{\mu_0/\mu}{2-\mu_0/\mu}\right)^2\frac{2 C_e^2-\frac{\kappa}{\lambda}C_\tau^2}{1+\frac{\kappa}{\lambda}+(\frac{\kappa}{\lambda}-1) \sec 2\theta}\,.\ee
For the MSSM parameters, we take $M_1=1$ TeV, $M_2=\mu_0 = 2$ TeV, $\mu = 500$ GeV, $\tan\beta = 2$ and choose $C_\tau=0.1$ as a specific example. The pair of the lightest neutralinos are dominantly Higgssinos with mass around 500 GeV in this case. With these, $V_{e N_3}$ and $V_{N_3 \tilde{h}_1^0}$ are obtained as functions of $r_e$, $m \xi$, $\kappa/\lambda$, $\theta$ and $M_{N_3}$. The first two can be extracted directly from the neutrino mass fits while the next two can be appropriately chosen from Fig. \ref{fig2} such that $N_3$ is the lightest RH neutrino. We chose $r_e$ and $m \xi$ from solution NH8 and randomly vary $\kappa/\lambda$ ($\theta$) in range $0.1$-$5$ ($1.0$-$2.0$) and compute $V_{e N_3}$ and $V_{N_3 \tilde{h}_1^0}$ as function of $M_{N_3}$.

The results are displayed in Fig. \ref{fig3}. We also show the typical value of $|V_{\alpha N}|^2$ expected naively from a seesaw relation, $|V_{\alpha N}|^2=\frac{0.05\, {\rm eV}}{M_{N}}$, as a solid line in the left panel of Fig. \ref{fig3}. The predicted values of $|V_{eN_3}|$ can be  larger than the typical expectation from seesaw estimates by several orders of magnitude. Also, a  number of points for $M_{N_3}<100$ GeV  fall either in the already excluded region or are quite close to it.

For the Higgssinos with mass around $500$ GeV that we have considered in the present analysis, their production cross section at a center-of-mass energy of $\sqrt{s} = 13$ TeV at the LHC is typically of ${\cal O}({\rm fb})$  \cite{Fiaschi:2023tkq}. This is notably smaller, by six to seven orders of magnitude, than the cross section for $W$-boson production. Nevertheless, the production of RH neutrinos resulting from the latter process is directly proportional to $|V_{\alpha N}|^2$. Consequently, RH neutrinos can be  predominantly generated through Higgsinos when the ratio $|V_{N_3 \tilde{h}_1^0}|^2/|V_{eN_3}|^2$ exceeds $10^7$. In Figure \ref{fig3}, we depict points satisfying $|V_{N_3 \tilde{h}_1^0}|^2/|V_{eN_3}|^2 > 10^8$ in a darker shade. These points stand out as particularly conducive to the production of RH neutrinos through Higgsinos. A dedicated analysis of the expected number of events would be needed to firmly establish the observability or otherwise of the RH neutrinos through the search of the displaced vertices at the LHC. We note however that a  similar analysis for two specific benchmark points is already done in \cite{Lavignac:2020yld} where  $|V_{N \tilde{\chi}_1^0}|^2 \sim  10^{-4}$ and $|V_{e N_3}|^2 \sim 10^{-12}$ are shown to lead to observable displaced vertices at the LHC for $M_N \sim 100$ GeV and for the similar values of $M_{1,2}$ and $\mu$. The magnitudes of $|V_{N_3 \tilde{h}_1^0}|$ and $|V_{eN_3}|$ obtained for $M_{N_3} \sim 100$ GeV in Fig. \ref{fig3} are comparable and may even lean towards being somewhat larger. Hence, the parameter range within the current model presents an opportunity to investigate RH neutrinos in the direct search experiments, particularly when their signatures through conventional processes are suppressed.

\section{Discussions}
\label{sec:disc}
Extension of the SM to include the LN as an explicit global or local symmetry is primarily aimed at providing tiny but non-vanishing masses for the light neutrinos through the seesaw mechanisms. On the other hand, extension with enlarged space-time symmetries such as supersymmetry is aimed to ameliorate some structural issues pertaining to the SM. In general, these are two separate classes of new physics and they may be operating at two different scales. They are often combined to play complementary roles. A more intriguing possibility is that  both these frameworks  arise from a common mechanism and, therefore, their scales are closely related. We have put forward one such mechanism and discussed its most relevant  phenomenological aspects in detail.

The mechanism uses a global LN symmetry and three SM singlet superfields, two of which are charged under the $U(1)_L$. The Lorentz scalar components of these superfields acquire non-zero vacuum expectation values and break $U(1)_L$ spontaneously only when the soft terms in supergravity are switched on. This relates the $U(1)_L$ breaking scale with the SUSY breaking scale which  is proportional to the gravitino mass $m_{3/2}$. The fermionic components of the SM singlet superfields which play the role of RH neutrinos also get masses of order $m_{3/2}$. The same scale also governs the neutrino-neutralino mixing through $R$-parity violation effectively induced by the VEVs of the SM singlet scalars. It is also possible to link the MSSM $\mu$-parameter with the common scale $m_{3/2}$ utilising the vacuum structure of the RH sneutrinos.

The identification of various scales with a common scale $m_{3/2}$ in the underlying framework leads to the following interesting and noteworthy observations:
\begin{itemize}
\item The light neutrinos acquire masses through the usual seesaw mechanism as well as from $R$-parity violation. The realistic neutrino mass spectrum requires that both these contributions are of similar magnitude. In the case of $\mu \sim m_{3/2}$ (i.e. conformal superpotential), both these contributions are proportional to $m_{3/2}^{-1}$. Therefore, the usual scale of the seesaw would also imply high-scale SUSY. On the other hand, keeping the SUSY scale close to the electroweak scale necessarily leads to low scale seesaw mechanism with RH neutrino masses of the order of a few hundred GeV.
\item If $\mu$ is kept independent of $m_{3/2}$ and held fixed close to the electroweak scale, the viable light neutrino masses do not allow a large hierarchy between $\mu$ and $m_{3/2}$. This again implies a low-scale seesaw mechanism and a SUSY breaking scale not very far from the electroweak scale. In particular, the gauginos and RH neutrinos are not allowed to be arbitrarily heavy in this case.
\item The spontaneous breaking of $U(1)_L$ at the scale $\sim\, m_{3/2}$ leads to a Majoron with coupling to the  light neutrinos of ${\cal O}(m_\nu/m_{3/2})$. This turns out to be small enough to satisfy all the present constraints even for massless Majoron.
\item The framework offers an interesting possibility of probing the existence of near weak-scale RH neutrinos through their displaced vertex signature in which they are produced dominantly from Higgssinos at the colliders and subsequently decay into the SM particles. It is seen that for a range of parameters, the model predicts sizeable mixings between Higgssino and RH neutrino and between the heavy and light neutrinos enhancing the possibility for direct detection.
\end{itemize}
All along, we have assumed that supergravity is responsible for the generation of the soft SUSY breaking. However, the underlying mechanism is applicable even if the SUSY breaking arises from some other sources.

The basic framework presented in this work utilizes an exact global LN symmetry which is spontaneously broken through soft SUSY breaking terms. The considered superpotential and the soft terms in the conformal case also possess a discrete $Z_3$ symmetry. It is known that the spontaneous breaking of these exact symmetries lead to formation of topological defects like comic strings and domain walls. However, small explicit breaking of such symmetries, possibly induced through gravity effects (see for example, \cite{Rothstein:1992rh,Akhmedov:1992hh,Akhmedov:1992hi}), can make these structures unstable. For example, spontaneous breaking of an approximate global $U(1)$ is shown to lead to strings connected by domain walls which decay quickly and do not dominate the universe \cite {Vilenkin:1982ks}. Similarly, the domain walls produced due to $Z_3$ breaking can be made to  disappear effectively before nucleosynthesis if Planck scale suppressed explicit $Z_3$ breaking terms are introduced \cite{Panagiotakopoulos:1998yw}. From a phenomenological standpoint, explicitly breaking LN results in a small mass for the Majoron the exact magnitude of which depends on the nature of the $Z_3$ breaking. While it could also contribute to light and heavy neutrino masses, the impact remains negligibly small with Planck scale suppressed LN symmetry breaking. Thus, small explicit breaking of the underlying global symmetries can address cosmological concerns without significantly altering the primary mechanism for neutrino mass generation presented in this study. The non-conformal case considered here does not have a discrete $Z_3$ symmetry and is thus free from the domain wall problem.

\section*{Acknowledgements}
This work is partially supported under the MATRICS project (MTR/2021/000049) by the Science \& Engineering Research Board (SERB), Department of Science and Technology (DST), Government of India. KMP acknowledges support from the ICTP through the Associates Programme (2023-2028) where part of this work was completed.

\appendix

\section{Potential minimization and VEVs of RH sneutrinos}
\label{app:minimization}
In this Appendix, we provide a derivation of the vacuum expectation values given in Eq. (\ref{N_vevs}). The relevant scalar potential of the model derived from $W_S$ in Eq. (\ref{WRH}) along with the soft terms is obtained as
\be \label{V}
V = |\lambda|^2 |\tilde{N}_\tau|^2 \left(|\tilde{N}_e|^2 + {|\tilde{N}_\mu|^2} \right) + |\lambda \tilde{N}_e \tilde{N}_\mu + \kappa \tilde{N}_\tau^2|^2 + V_{\rm soft}\,,\ee
where $V_{\rm soft}$ is given in Eq. (\ref{soft1}).

We parametrize the VEVs of the three scalar fields as
\be \label{app:vev1}
\langle \tilde{N}_\tau \rangle = q\,,~~\langle \tilde{N}_e \rangle = p \cos\phi\,,~~ \langle \tilde{N}_\mu \rangle = p \sin\phi\,.\ee
The parameters $p$, $q$ and $\phi$ are to be obtained by solving the minimization conditions $\frac{\partial V}{\partial \tilde{N}_\alpha}=0$ at the minimum for $\tilde{N}_\alpha = \tilde{N}_\tau, \tilde{N}_e, \tilde{N}_\mu$. For simplicity, we assume all the parameters in $V$ real and also require real solutions for the VEVs. Using a linear combination of two of the minimization conditions
\be \label{cond_comb1}
\cos\phi\, \frac{\partial V}{\partial \tilde{N}_e} - \sin\phi\,\frac{\partial V}{\partial \tilde{N}_\mu} =0\,,\ee
we obtain
\be \label{cos2th}
\cos 2\phi = \frac{(a_{N_\mu}^2 - a_{N_e}^2)\, m_{3/2}^2}{(a_{N_\mu}^2 + a_{N_e}^2)\, m_{3/2}^2 + 2 q^2 \lambda^2}\,,\ee
at the minimum of the potential. Using the orthogonal linear combination
\be \label{cond_comb2}
\sin\phi\, \frac{\partial V}{\partial \tilde{N}_e} + \cos\phi\,\frac{\partial V}{\partial \tilde{N}_\mu} =0\,,\ee
we solve for $p^2$ and obtain
\be \label{p2}
p^2=0\,~{\rm or}\,~p^2 = -\frac{1}{\lambda^2}\left((a_{N_e}^2+a_{N_\mu}^2) m_{3/2}^2 + 2 q^2 \lambda^2 + \frac{2q\lambda}{\sin 2\phi} (m_{3/2} A_\lambda + q \kappa)\right)\,.\ee

Next, we assume $a_{N_e}^2 = a_{N_\mu}^2$ for simplification which leads to $\cos 2\phi = 0$ and degenerate VEVs for $\tilde{N}_e$ and $\tilde{N}_\mu$. Using the remaining minimization condition, $\frac{\partial V}{\partial \tilde{N}_\tau}=0$, and substituting the non-vanishing solution for $p^2$ given in Eq. (\ref{p2}) along with $\cos 2\phi = 0$, we find the following equation:
\be \label{cond_q}
q^3+b q^2 + c q + d = 0\,,\ee
with
\beqa \label{bcd}
b & = & \frac{3 A_\lambda (\kappa + \lambda) - A_\kappa \kappa}{2 \lambda (2 \kappa + \lambda)}\,m_{3/2}\,, \nonumber \\
c & = & \frac{(A_\lambda^2 - a_{N_\tau}^2)\lambda + 2 a_{N_e}^2 (\lambda + \kappa)}{2 \lambda^2 (2 \kappa + \lambda)}\,m_{3/2}^2\,, \nonumber \\
d & = & \frac{A_\lambda a_{N_e}^2}{2 \lambda^2 (2 \kappa + \lambda)}\,m_{3/2}^3\,.  \eeqa
The other solution, $p^2=0$, leads to $q=0$. Eq. (\ref{cond_q}) needs to be solved to determine $q$ which then can be used to find the VEVs in terms of the parameters of the scalar potential.

It is convenient to convert Eq. (\ref{cond_q}) into the so-called ``Depressed cubic" equation \cite{DicksonFirstCI} by the following change of variable:
\be \label{variable}
q = u - \frac{b}{3}\,. \ee
In the new variable, the equation has no quadratic term and it is given by
\be \label{cond_u}
u^3 + P u + Q = 0\,,\ee
where  
\beqa \label{PQ}
P & = & \frac{3 c -b^2}{3}\,, \nonumber \\
Q & = & \frac{2b^3 - 9 bc + 27 d}{27}\,.  \eeqa
The discriminant of the Depressed cubic equation is given by
\be \label{Delta}
\Delta = -(4 P^3 + 27 Q^2)\,. \ee

The following possibilities exist for the solutions \cite{DicksonFirstCI} of Eq. (\ref{cond_u}):
\begin{itemize}
\item If $\Delta > 0$, then the underlying polynomial has three distinct real roots. They are explicitly given by
\be \label{real_root}
u_k = 2 \sqrt{\frac{-P}{3}}\, \cos\left[\frac{1}{3} \cos^{-1}\left(\frac{3Q}{2P}\sqrt{\frac{-3}{P}} \right)-\frac{2 \pi k}{3}\right]\,,~~~k=0,1,2\,. \ee
\item For $\Delta = 0$ and $P \neq 0$, the polynomial has three real roots out of which two are degenerate. These are
\be \label{roots_D0}
u_1 = \frac{3 Q}{P}\,,~~~~u_{2,3} = - \frac{3 Q}{2 P}\,. \ee 
\item If $\Delta < 0$ then the polynomial has one real and two complex roots. The real root can be represented, for $P < 0$, as
\be \label{real_root1}
u_0 = -2 \frac{|Q|}{Q}\,\sqrt{-\frac{P}{3}}\,\cosh\left[\frac{1}{3} \cosh^{-1} \left(-\frac{3|Q|}{2P}\sqrt{-\frac{3}{P}} \right) \right]\,,\ee
and, for $P > 0$, as
\be \label{real_root2}
u_0 = -2 \sqrt{\frac{P}{3}}\,\sinh\left[\frac{1}{3} \sinh^{-1}\left(\frac{3Q}{2P}\sqrt{\frac{3}{P}} \right) \right]\,.\ee
\end{itemize}
Substituting $b$, $c$ and $d$ from Eq. (\ref{bcd}) in Eq. (\ref{PQ}), one can determine $\Delta$ which turns out to be proportional to $(m_{3/2}/\lambda)^6$. With an appropriate choice of the values of the soft parameters, $\Delta \ge 0$ can be obtained leading to the real solutions for $q_i$.

As an explicit example, consider 
\be \label{choice_soft}
A_\kappa = A_\lambda \equiv A\,,~~ a_{N_e}^2 = a_{N_\tau}^2 = A^2/8\,. \ee
This leads to $\Delta = 0$ for which the roots can be estimated from Eq. (\ref{roots_D0}). Substituting this choice of the soft parameters in Eq. (\ref{bcd}) and using Eq. (\ref{roots_D0}) and (\ref{variable}), we find the real solutions $q_i$ as
\be \label{q_sol}
q_1 = -\frac{A m_{3/2}}{2 \kappa + \lambda}\,~~~{\rm and}~~~q_{2} = q_3 = -\frac{A m_{3/2}}{4 \lambda}\,\ee
Substitution of the above in $p^2$ expression Eq. (\ref{p2}) leads to
\be \label{p_sol}
p_1^2 = -\frac{A^2 m_{3/2}^2 (\lambda-2\kappa)^2}{4 \lambda^2 (\lambda+ 2\kappa)^2}\,~~~{\rm and}~~~p_{2}^2 = p_3^2 = \frac{A^2 m_{3/2}^2 (\lambda- \kappa)}{8 \lambda^3}\,.\ee
The first solution leads to an imaginary $p$ while the second corresponds to real $p$ if $\lambda>\kappa$. Moreover, the solution corresponding to $(q_{2},p_{2})$ is a global minimum.

In summary, for $a_{N_\mu} = a_{N_e}$, the scalar potential given in Eq. (\ref{V}) can lead to the following VEV configuration:
\be \label{VEVs_sol}
\langle \tilde{N}_\tau \rangle = \frac{C_\tau}{\lambda} m_{3/2}\,,~~\langle \tilde{N}_e \rangle = \langle \tilde{N}_\mu \rangle = \frac{C_e}{\lambda} m_{3/2}\,,\ee
where $C_\tau$ and $C_e$ are dimensionless parameters determined by $\lambda$, $\kappa$ and the soft parameters. The non-trivial relation between the VEVs obtained in Eq. (\ref{p2}) implies the following correlation between $C_e$ and $C_\tau$:
\be \label{Ce_Ctau_gen}
\frac{C_e^2}{C_\tau^2} = - 1 - \frac{\kappa}{\lambda} - \frac{a_{N_e}^2}{C_\tau^2} - \frac{A_\lambda}{C_\tau}\,. \ee
For the specific choice of the soft terms used in Eq. (\ref{choice_soft}) and the resulting $q_2$ in Eq. (\ref{q_sol}), the above becomes
\be \label{Ce_Ctau}
\frac{C_e^2}{C_\tau^2} = 1 - \frac{\kappa}{\lambda} \,, \ee
which is completely determined in terms of the ratio $\kappa/\lambda$. However, the desired value of $C_\tau$ and $C_e$ can be obtained by the appropriate choice of $a_{N_e}$ and $A_\lambda$ in general. Therefore, the VEVs in Eq. (\ref{VEVs_sol}) are parametrized in a general way and considered for phenomenological analysis.

\section{Calculation of $k_\alpha$}
\label{app:k_calc}
We derive expressions for $k_\alpha$ displayed in Eqs. (\ref{kalpha1},\ref{kalpha2}) in this Appendix following a similar calculation carried out earlier in \cite{Giudice:1992jg,Joshipura:2002fc} for a different framework. When the scalar components of $\hat{N}_{e,\tau}$ take VEVs, the bilinear terms in $W$ are given by
\be \label{W_bl}
W \supset -\, \epsilon_\alpha\, \hat{L}^\prime_\alpha \hat{H}_2\, -\, \mu\, \hat{H}^\prime_1 \hat{H}_2\, \ee
Here, $\epsilon_\alpha = - \lambda_\alpha \langle \tilde{N}_e \rangle$ and we consider a general case in which $\mu$ contains both the contributions such as shown in Eq. (\ref{newmu}). The scalar potential evaluated from the above is then given by
\beqa \label{VFD_app}
V_{F,D} &=& \left|\mu \tilde{H}^\prime_1 + \epsilon_\alpha \tilde{L}^\prime_\alpha \right|^2 + \left(|\mu|^2+|\epsilon_\alpha|^2 \right) |\tilde{H}_2|^2 + |\eta|^2 |\tilde{H}_1^\prime|^2 |\tilde{H}_2|^2 \nonumber \\
&+& \frac{1}{8}(g^2 + g^{\prime 2})\left(|\tilde{H}_1|^2-|\tilde{H}_2|^2 + |\tilde{L}^\prime_\alpha|^2 \right)^2\,,
\eeqa
where the last term in the first line arises from the $F$-term of $\hat{N}_\tau$. Further, the relevant soft terms can be written as
\beqa \label{vsoft_app}
V_{\rm soft} &=& m_{3/2} \left(A_\alpha \epsilon_\alpha \tilde{L}^\prime_\alpha \tilde{H}_2  + B_\mu \mu  \tilde{H}^\prime_1 \tilde{H}_2 + {\rm c.c.} \right) \, \nonumber \\
&+& m_{3/2}^2 \left(a_{H_1}^2\, |\tilde{H}^\prime_1|^2 + a_{H_2}^2\, |\tilde{H}_2|^2 + a_{L _\alpha}^2\, |\tilde{L}^\prime_\alpha|^2 \right) \,,\eeqa
where we have defined
\be \label{Bmu_def}
B_\mu \mu  =  B_{\mu_0} \mu_0 - A_\eta \eta \langle \tilde{N}_\tau \rangle\,.  \ee
In Eqs. (\ref{VFD_app},\ref{vsoft_app}), the index $\alpha$ is summed over.

Rotating away the term proportional to $\epsilon_\alpha$ form Eq. (\ref{W_bl}) using the redefinitions:
\be \label{Rbasis}
\mu \hat{H}^\prime_1+ \epsilon_\alpha \hat{L}_\alpha^\prime = \mu \hat{H}_1\,,~~-\epsilon_\alpha \hat{H}^\prime_1 + \mu \hat{L}^\prime_\alpha = \mu \hat{L}_\alpha\,,\ee
one finds that $V_{F,D}$ and $V_{\rm soft} $ get modified to
\beqa \label{new_V_app}
\tilde{V}_{F,D} &=&  |\mu |^2 \left(|\tilde{H}_1|^2 + |\tilde{H}_2|^2\right)  + |\eta|^2 \left|\tilde{H}_1 -\frac{\epsilon_\alpha}{\mu} \tilde{L}_\alpha \right|^2 |\tilde{H}_2|^2 \nonumber \\
&+& \frac{1}{8}(g^2 + g^{\prime 2})\left(|\tilde{H}_1|^2-|\tilde{H}_2|^2 + |\tilde{L}_\alpha|^2 \right)^2 \,, \nonumber \\
\tilde{V}_{\rm soft} &=& m_{3/2} \left((A_\alpha -B_\mu ) \epsilon_\alpha \tilde{L}_\alpha \tilde{H}_2  + B_\mu \mu  \tilde{H}_1 \tilde{H}_2 + {\rm c.c.} \right) \, \nonumber \\
&+& m_{3/2}^2 \left(a_{H_1}^2\, |\tilde{H}_1|^2 + a_{H_2}^2\, |\tilde{H}_2|^2 + a_{L _\alpha}^2\, |\tilde{L}_\alpha|^2 - \left\{\frac{\epsilon_\alpha}{\mu} (a_{H_1}^2 - a_{L_\alpha}^2)\, \tilde{H}_1^* \tilde{L}_\alpha + {\rm c.c.} \right\} \right)\,.\eeqa
In the new basis, the VEV of the electrically neutral components of the $\tilde{L}_\alpha$ are denoted by $\omega_\alpha$. It is related to the VEV $\omega^\prime_\alpha$ in the original basis by  Eq. (\ref{Rbasis}).
Using the defining relation, $\omega_\alpha = k_\alpha \epsilon_\alpha$, and minimizing the potential $\tilde{V}_{F,D}+\tilde{V}_{\rm soft}$, we obtain
\be \label{k_calculated_0}
k_\alpha \simeq \frac{v_1}{\mu}\frac{\left(a_{H_1}^2 - a_{L_\alpha}^2  + |\eta|^2 \frac{v_2^2}{m_{3/2}^2}\right) + \frac{\mu}{m_{3/2}} (B_\mu - A_\alpha) \tan\beta }{a_{L_\alpha}^2 + \frac{1}{2} \frac{M_Z^2}{m_{3/2}^2}\, \cos2\beta}\,.\ee

Neglecting the contributions of ${\cal O}(v^2/m_{3/2}^2)$, the above can be further simplified to
\be \label{k_calculated}
k_\alpha \simeq \frac{v_1}{\mu} \left(\frac{a_{H_1}^2 - a_{L_\alpha}^2}{a_{L_\alpha}^2} \right) + \frac{v_1}{m_{3/2}} \left( \frac{B_\mu - A_\alpha}{a_{L_\alpha}^2}\right) \tan\beta \,.\ee
The first term in $k_\alpha$ is suppressed for the universal soft masses. Depending on the presence or absence of bare $\mu_0$ in the model, the nature of the second term in $k_\alpha$ can be decided. For example, when $\mu_0 = B_{\mu_0} = 0$, using Eq. (\ref{Bmu_def}) and the fact that $\mu = -\eta \langle \tilde{N}_\tau \rangle$,   Eq. (\ref{k_calculated}) can be reduced to
\be \label{k_calculated_zeromu}
k_\alpha \simeq \frac{v_1}{\mu} \left(\frac{a_{H_1}^2 - a_{L_\alpha}^2}{a_{L_\alpha}^2} \right) + \frac{v_1}{m_{3/2}} \left( \frac{A_\eta - A_\alpha}{a_{L_\alpha}^2}\right) \tan\beta \,.\ee
Both terms in $k_\alpha$ can then be suppressed for the universal soft terms.

\bibliography{references}

\end{document}